\documentclass[pra,twocolumn,showpacs,superscriptaddress]{revtex4-1}
\usepackage[english]{babel}
\usepackage[dvips]{graphicx}
\usepackage{color}
\usepackage{latexsym}
\usepackage{amsmath}
\usepackage{amsthm}
\usepackage{amsfonts}
\usepackage{amssymb}
\usepackage{mathtools}
\usepackage{array}
\usepackage{cancel}
\usepackage{algorithmic}
\usepackage[dvipsnames]{xcolor}
\usepackage{svg}
\usepackage{hyperref}
\hypersetup{
    colorlinks=true, 
    linkcolor=blue, 
    citecolor=blue,
    urlcolor=black, 
    }
\usepackage{orcidlink}

\newcolumntype{P}[1]{>{\centering\arraybackslash}p{#1}}

\begin{document}

\pagenumbering{arabic}

\title{Anomalous Dissipation in Current Biased Josephson Systems}
\author{Johannes Hauff}
\affiliation{Institut f\"ur komplexe Quantensysteme,  Universit{\"a}t Ulm and IQST, D-89069 Ulm, Germany}
\author{Niklas Gaiser\,\orcidlink{0000-0002-8539-0292}}
\affiliation{Institut f\"ur komplexe Quantensysteme,  Universit{\"a}t Ulm and IQST, D-89069 Ulm, Germany}
\author{Joachim Ankerhold\,\orcidlink{0000-0002-6510-659X}}
\affiliation{Institut f\"ur komplexe Quantensysteme,  Universit{\"a}t Ulm and IQST, D-89069 Ulm, Germany}
\author{Dominik Maile\,\orcidlink{0000-0002-0740-971X}}
\affiliation{Institut f\"ur komplexe Quantensysteme,  Universit{\"a}t Ulm and IQST, D-89069 Ulm, Germany}
\begin{abstract}
A new phase diffusive regime in a current biased Josephson junction is theoretically explored which originates from embedding the junction in a circuit environment with anomalous dissipation. This is realized by placing parallel to the junction a resistor in series with a capacitor such that  electromagnetic fluctuations  effectively couple also to the charge of the junction. This leads to rich Josephson dynamics, in particular for the switching of the junction out of a zero voltage state.
Modelled as the escape process of a fictitious phase-particle out of a metastable well, a detailed study  reveals  that anomalous dissipation has a strong impact at low temperatures when quantum tunneling dominates against thermal activation. As a manifestation, a  regime is found, where for realistic circuit parameters the quantum escape process is substantially enhanced, followed by a short voltage pulse and re-trapping with high probability. This class of circuits may be leveraged for detecting microwave photons or dissipative quantum annealing processes. In addition, the analysis provides a general framework for engineering dissipative dynamics in nonlinear systems using anomalous environments. 
\end{abstract}
\date{\today}
\maketitle

%
%
%
%
\section{Introduction}
Superconducting quantum circuits are among the most versatile platforms for current and future quantum technologies \cite{zagoskin_quantum_2011}.
The basic element is a Josephson junction (JJ) which, due to its embedding in an electrical circuit, is inevitably subject to fluctuations of
and energy exchange with the electromagnetic environment \cite{devoret_measurements_1985,Schon:1990kj,Martinis1990}. This in turn gives rise to a wealth of dynamical regimes for the two conjugate variables of the junction, the phase and the charge, from the classical down to the deep quantum domain. From a technological point of view, circuit design allows to address specific modes of operation for example, to encode  superconducting quantum bits  \cite{Wendin2017,Krantz2019} or to detect weak magnetic fields \cite{clarke2006}.  

Corresponding progress has remained strongly tied to the theoretical understanding of open quantum systems \cite{Weiss2012,Golubev2019}.
In fact, already in the first seminal experimental realization \cite{devoret_measurements_1985}, the switching of  a current biased Josephson system out of the zero voltage state could effectively be described as the quantum dissipative dynamics of a fictitious phase-particle in a tilted washboard potential. 
\begin{figure}[t]
	\centering
\includegraphics[scale=0.017]{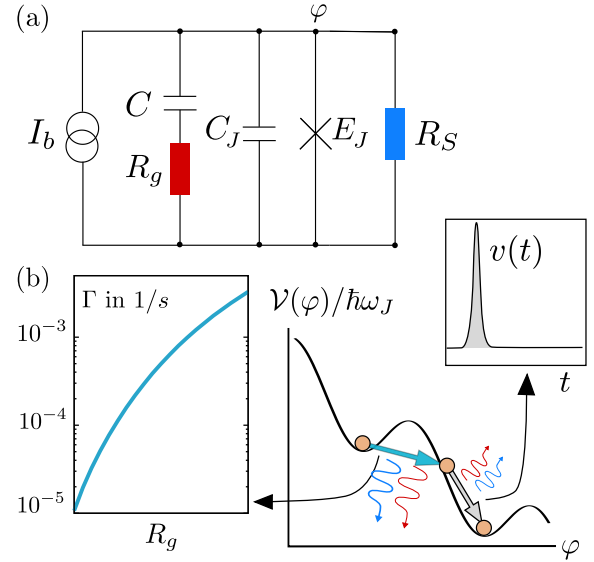}
	\caption{(a) Current-biased Josephson junction,
connected in parallel to a shunt resistor $R_S$ and 
	and to a branch containing an external capacitor $C$ and a resistor $R_g$.
	(b) Sketch of the tilted washboard potential with the main processes analysed in the article. The fictitious phase particle is initially trapped in a metastable potential minimum. The decay rate out of this minimum $\Gamma$ is enhanced by increasing $R_g$. The dynamics after the tunneling event yield a voltage pulse stemming from acceleration due to $R_g$ followed by a re-trapping behaviour where $R_g$ and $R_S$ both damp the dynamics.   }
	\label{Fig.1}
\end{figure}
Recently, this setting has regained new interest as a promising platform for single photon detection in the microwave regime \cite{Golubev2021,Gatti2021,Revin2020,Krasnov2024}.

The most common modelling of the classical Josephson dynamics is the so-called Resistively and Capacitively Shunted Junction (RCSJ) model, where a semi-infinite transmission line couples to the Josephson phase, effectively yielding Ohmic dissipation.
The general theory behind this approach was formulated by Caldeira and Leggett \cite{caldeira_quantum_1983,caldeira_quantum_1984} in form of system-reservoir models, where a system degree of freedom interacts bilinearly with a quasi-continuum of harmonic modes. In the classical domain, this leads to a generalized Langevin equation for the system degree of freedom with dissipation and fluctuations being related by a fluctuation-dissipation theorem. 

This framework also allows to consider a minimal extension of the RCSJ model with an additional resistor $R_g$ in series to a capacitor $C$ placed in parallel to the JJ (see Fig.~\ref{Fig.1}a).
The effect of such an environment has already been studied both theoretically and experimentally  \cite{Martinis1990,Vion1996,Joyez1999,Falci2003,Clarke2005,Zorin2005,Hassel2006,Fistul2007,Gatti2021,Kaur2021}.  So far, the most complete classical analysis has been given in \cite{Martinis1990}, explaining the impact of the $C-R_g$ shunt on the hysteresis of a current biased JJ. Experimentally, this design has been used in several contexts, for example, to improve qubit readout \cite{Clarke2005}, to analyse measured switching current distributions \cite{Gatti2021}, and to describe switching rates in high-TC superconducting junctions \cite{Ustinov2006}.

With respect to the switching of the JJ out of a zero voltage state, the simplest picture considers  the dynamics of the phase in a tilted $cos$-potential (washboard potential), where thermal fluctuations lead to an Arrhenius type temperature dependence of the escape rate from a metastable state (with weak dependence on the damping strength). This process of thermal activation is followed by the subsequent downhill-running of the phase-particle or its re-trapping in one of the subsequent potential wells, depending on the energy loss (dissipation strength).  However, this type of description applies only to the high temperature regime. At lower temperatures, quantum tunneling tends to dominate and exhibits  only very weak temperatures dependence but strong dependence on the dissipation. Generically, in a circuit with Ohmic shunt resistor $R_S$ (see Fig.~\ref{Fig.1}a), i.e. conventional Ohmic friction, the probability of quantum tunneling is suppressed \cite{caldeira_quantum_1983,HanggiReview1990,Weiss2012} and the so-called crossover temperature between thermal activation  and quantum tunneling is reduced.  

Surprisingly, for the switching process,  the impact of an additional resistor $R_g$ as in Fig.~\ref{Fig.1}a is much less explored. Due to the capacitor in series with $R_g$, the environment effectively couples to the {\em charge}  of the system. Since for the JJ, the phase and the charge are canonically conjugate variables, the effect of this type of dissipative coupling is analogous to \textit{anomalous} dissipation, as termed by Leggett \cite{leggett_quantum_1984} and explored as \textit{momentum }dissipation in several other works over the last decade \cite{Pollak2006,ankerhold_dissipation_2007,maile_effects_2020,maile_exponential_2021,kohler_dissipative_2006,cuccoli_quantum_2001,rastelli_dissipation-induced_2016,maile_quantum_2018}. 
In the quantum regime, qualitatively, position and momentum dissipation can be understood in the context of the measurement process: In both cases, the environment acts as a measurement apparatus squeezing either phase (position) or charge (momentum) so that, as Heisenberg's uncertainty relation dictates, the respective conjugate variable shows enhanced quantum fluctuations.  
For example, for a phase qubit such a scenario leads to an increase of the tunneling rate of the phase particle through the barrier and therefore, counter-intuitively, to an enhancement of a quantum effect, as recently theoretically discussed in \cite{Maile2022}.

In this article, we provide a comprehensive analysis of the circuit displayed in Fig.~\ref{Fig.1}. It can be seen as a generalization of the RCSJ model with an additional resistor $R_g$ in series with the capacitor. For this setting, we aim to address three main questions: How does $R_g$ in combination with $R_S$ affect - 1) the dynamics of a non-linear system? - 2) the decay rate out of a metastable state in the presence of temperature? - 3) the most probable exit point of the decay process? By answering these three questions, we obtain a detailed picture of the dissipative quantum dynamics of this quantum circuit with particular focus on the previously unexplored regime of strong bias $I_B/I_C >0.9$ at strong coupling to the anomalous environment. There, we uncover a new phase diffusive regime, highlighting new reservoir engineering possibilities in circuitQED.  

First, in Sec.~\ref{Sec.2}, we calculate the quantum Langevin equation for the genuine effect of $R_g$  coupled to an arbitrary electrical circuit and prove its equivalence to anomalous dissipation. Then, in the semiclassical regime,  the current biased JJ in the presence of $R_g$ and $R_S$ is studied and it is shown that $R_g$ introduces state-dependent \textit{damping} or \textit{amplification}, leading to interesting dynamical regimes for the Josephson phase. Specifically, we find that the voltage oscillations in the running state in the washboard potential are strongly amplified by $R_g$. We discuss the regime where the phase particle is deterministically re-trapped in the next minimum after an escape process and the impact of fluctuations on this effect. \\ 
\indent 
In Sec.~\ref{Sec.3}, we discuss the theoretical techniques used to solve the problem of a metastable decay from a local potential minimum in the presence of dissipation.  Corresponding results are shown in Sec.~\ref{Sec.meta}; they go beyond previous studies by discussing the impact of $R_g$ and $R_S$ on the decay rate of the metastable superconducting state including the effects of temperature. We identify that by increasing $R_g$, the quantum decay rate increases and that the crossover from escape dominated by quantum tunneling to the regime of thermal hopping appears at higher temperatures. Hence, the particle is predominantly tunneling with an enhanced rate over a wide range of temperatures.
To make contact to actual experiments, realistic parameters are used to provide specific theoretical predictions.\\ \indent 
In this context, we also discuss the most probable escape points of the decay process as a function of temperature as well as $R_g$ and $R_S$. We find that by increasing $R_g$ the phase is predominantly tunneling deep in the well, thus yielding a low escape point beyond the barrier. 
Combining the semiclassical description of the running state with the quantum description of the metastable decay, we provide a complete picture of this new dynamical domain of the phase in a washboard potential in the presence of general Ohmic environments and at strong coupling.
This way, we identify parameters where the tunneling rate is strongly enhanced,  {\em and} the classical dynamics of the phase particle in the running state leads to re-trapping in the next minimum of the washboard potential, termed the phase diffusive regime for classical JJs \cite{Vion1996,MELNIKOV1991,Kivioja2005,Haxell2023}.
By engineering the dissipative resistors, this dynamics can even be achieved for bias currents $I_b \sim I_C$, where $I_C$ is the critical current of the junction. Here, re-trapping occurs  already deterministically and is therefore different from re-trapping through thermal or quantum fluctuations as often discussed in the regime of strong bias \cite{MELNIKOV1991,Belzig2015}. 
Interestingly, after the switching process, the anomalous environment also leads to an amplification of the voltage pulse before re-trapping of the phase particle as sketched in Fig.~\ref{Fig.1}b.

Although the core of this work is dedicated to  the system displayed  in Fig.~\ref{Fig.1}, our results can be used to obtain a more general picture of the dissipative dynamics in arbitrary potential landscapes. We comment on this in Sec.~\ref{Sec.Outlook} and conclude in Sec.~\ref{Sec.Conclusion}.

\begin{figure}[t]
	\centering
\includegraphics[scale=0.02]{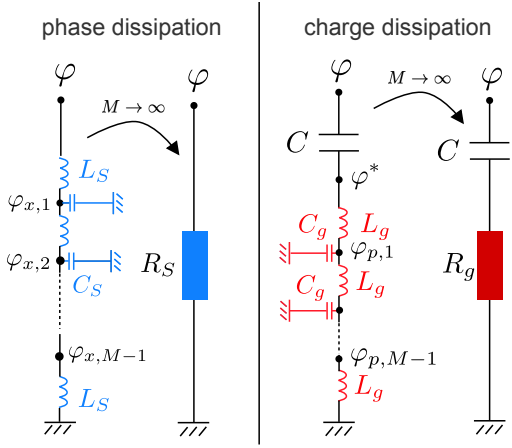}
	\caption{Two modes of dissipation in an electrical circuit described by a semi-infinite transmission line. \textbf{Left:}   The line is coupled to the phase $\varphi$. \textbf{Right:} The line is coupled to the phase ${\varphi}_*$ while the capacitor $C$ yields a coupling between $\dot\varphi$ and $\dot{\varphi}_*$. Both settings converge to Ohmic reservoirs in the limit $M\rightarrow \infty$. }
	\label{Fig.lines}
\end{figure}

\section{Langevin equations for conventional and anomalous dissipation}  
\label{Sec.2}
\subsection{Model Hamiltonians and quantum Langevin equations}
We start with the general system Hamiltonian $\hat{\mathcal{H}}_{S}
=\hat{Q}^2/(2C_{tot})
+ \mathcal{V}(\hat\varphi)$ 
with phase degree of freedom $\hat\varphi$ in a potential $\mathcal{V}(\hat{\varphi})$, where  the charge $\hat{Q}$  
is its conjugate operator (i.e. $ [\hat{\varphi},\hat{Q}]=i 2e $) and $C_{tot}$ is the total capacitance. 
The microscopic description of the dissipative resistors is formulated using transmission lines of semi-infinite length \cite{Devoret2017}. The total junction+environment Hamiltonian can then be expressed as 
\begin{equation}
\hat{\mathcal{H}}=\hat{\mathcal{H}}_{S}+\hat{\mathcal{H}}^{(\varphi)}_{IB}+\hat{\mathcal{H}}^{(Q)}_{IB}\,.
\end{equation}
Before we start to discuss the full model, we consider the transmission lines individually used to describe the resistors  $R_S$ and $R_g$ (see Fig.~\ref{Fig.lines}), respectively. 

One first finds  for the coupling to the phase $\hat\varphi$
\begin{align}
\label{eq:H_Bx}
\hat{\mathcal{H}}^{(\varphi)}_{IB}&=\frac{1}{2}\frac{\Phi_{0}^{2}}{L_{s}}\left(\hat{\varphi}-\hat{\varphi}_{x,1}\right)^{2} \nonumber \\ &+\sum_{n=1}^{M-1}\left(\frac{\hat{Q}_{x,n}^{2}}{2C_{s}}+\frac{1}{2}\frac{\Phi_{0}^{2}}{L_{s}}\left(\hat{\varphi}_{x,n}-\hat{\varphi}_{x,n+1}\right)^{2}\right),
\end{align}
with $M$ harmonic $LC$ oscillators and where $\Phi_0=\hbar/(2e)$. Equation.~(\ref{eq:H_Bx}) amounts to be the well studied \textit{Rubin model} \cite{Weiss2012}.  By calculating the Heisenberg equations of motion of the  Hamiltonian $\mathcal{H}^{(\varphi)}=\hat{\mathcal{H}}_{S}+\hat{\mathcal{H}}^{(\varphi)}_{IB}$ alone, we find by taking the limit $M\rightarrow \infty$ the well known Quantum Langevin equation
\begin{align}
\frac{\hbar}{2e}\ddot{\hat{\varphi}}(t)=&-\frac{1}{C_{tot}}\mathcal{V}^{'}(\hat\varphi(t))  - \gamma(t)\frac{\hbar}{2e}\hat\varphi (0)\label{Eq.langevinphs} \\ &-\int_{0}^{t}du\,{\gamma}(t-u)\frac{\hbar}{2e}\dot{\hat\varphi}(u)+\frac{\hat{\boldsymbol{I}}_{\varphi}(t)}{C_{tot}}, \nonumber
\end{align}
with $\gamma(t)=1/(C_{tot}R_S)\omega_R e^{-\omega_Rt}$. Here, $\omega_R$ is a Drude high frequency cutoff  of the resistor $R_S$ we choose phenomenologically, and $\hat{\boldsymbol{I}}_\varphi(t)$ is a fluctuating current stemming from the resistor $R_S$. The second term is the so called initial slippage, appearing because we have not specified that the bath and the system are in equilibrium at $t=0$, yet.

Now, we turn to charge dissipation and make again use of a transmission line in the form of Eq.~(\ref{eq:H_Bx}). This is  coupled via an additional capacitance $C$ to the charge $\hat{Q}$ of the circuit which leads to the Hamiltonian 
\begin{align}
\hat{\mathcal{H}}^{(Q)}_{IB}&=\frac{C}{2C_{J}C_{tot}}\left(\hat{Q}+\frac{C_{tot}}{C}\hat{Q}_{*}\right)^{2}\!\!+\frac{1}{2}\frac{\Phi^2_{0}}{L_{g}}\left(\hat{\varphi}_{*}-\hat{\varphi}_{p,1}\right)^{2} \nonumber  \\ &+\sum_{n=1}^{M-1}\left(\frac{\hat{Q}_{p,n}^{2}}{2C_{g}}+\frac{1}{2}\frac{\Phi^2_{0}}{L_{g}}\left(\hat{\varphi}_{p,n}-\hat{\varphi}_{p,n+1}\right)^{2}\right), \label{Eq.QHam}
\end{align}
where the corresponding charges and phases are also displayed in Fig.~\ref{Fig.lines}. Eq.~(2) and (3) highlight the important difference between the two dissipative couplings. Particularly, and contrary to the velocity coupled Hamiltonian given in \cite{Ford1988}, it is not possible to convert one into the other via an unitary transformation.
To exemplify this further, we also take the continuum limit $M\rightarrow \infty$ and use  the Hamiltonian $\mathcal{H}^{(Q)}=\mathcal{H}_S+\mathcal{H}^{(Q)}_{IB}$ to calculate the Langevin equation for the coupling to the charge, i.e. 
\begin{align}
\frac{\hbar}{2e}\ddot{\hat{\varphi}}(t)=&-\frac{1}{C_{tot}}\mathcal{V}^{'}(\hat\varphi(t)) - \frac{C}{C_{tot}^2}\,\eta(t)\left({\hat{I}}_0+{\hat{I}}_g\right) \label{Eq.eqofmo} \\ &-\frac{C}{C^2_{tot}}\int_{0}^{t}du\,{\eta}(t-u)\frac{d}{du}\left(\mathcal{V}^{'}(\hat\varphi(u))\right)+\frac{\hat{\boldsymbol{I}}_{Q}(t)}{C_{tot}}, \nonumber
\end{align}
 where 
\begin{align}
{\eta}(t)=CR_{g}\tilde\omega_{c}\,\exp({-t\tilde\omega_{c}}) \quad \text{with} \quad \tilde{\omega}_c=\frac{\omega_{R}}{1+\frac{CC_{J}}{C_{tot}}R_{g}\omega_{R}}
\end{align}
is a (Ohmic) damping kernel
with a characteristic high frequency cutoff $\tilde{\omega}_c$, depending on the circuit parameters and $\omega_R$ (the high frequency cutoff of the spectral density describing $R_g$). 
Furthermore, here the initial slippage term is formed by the second term containing the current operator through the junction $\hat{I}_0= \dot{\hat{Q}}(0) $ and $\hat{I}_g = -\tilde{\omega}_C \hat{Q}(0) $  encoding the current through the $C-R_g$ branch, at time $t=0$, respectively.  We  also defined $\hat{\boldsymbol{I}}_{Q}(t)$ containing all initial values from the bath and therefore, the thermal and quantum fluctuations stemming from $R_g$.

In Eq.~(\ref{Eq.eqofmo}), the time retarded kernel $\eta(t)$ couples to the second time derivative of the charge $\ddot{\hat{Q}}(t) = -\frac{d}{dt} \,\mathcal{V}'(\varphi(t))$ and not - as in the case of a shunt resistor - to the velocity $\dot{\hat\varphi}$ (see Eq.~(\ref{Eq.langevinphs})). This Langevin equation can also be obtained by using the admittance $Z^{-1}_{s}(\omega)=i \omega C/\left(1+i \omega R_{g}C\right)$ within a usual Caldeira Leggett description or by applying Kirchhoff rules to the currents in the circuit.
However, by explicitly eliminating the bath degrees of freedom the form (\ref{Eq.eqofmo}) and particularly the fluctuating current $\hat{\boldsymbol{I}}_{Q}$ naturally arises and highlights that the resistor $R_g$ creates a dissipative coupling that is  in complete analogy to the Langevin equation for momentum dissipation discussed in \cite{cuccoli_quantum_2001}, providing a proof of the relation between the circuit described by Eq.~(\ref{Eq.QHam}) (see also Fig.~\ref{Fig.1}) and the more abstract concept of anomalous (or momentum) dissipation. However, note that while in the circuit discussed here the cutoff frequency $\tilde{\omega}_c$ depends on the circuit parameters and particularly also on $R_g$, for general momentum dissipation the spectral density and its cutoff frequency is typically  chosen on demand. Below, we will show in which regime the kernel $\eta(t)$ can be approximated by a time-local response.

Although the derived quantum Langevin equations contain quantum operators, the direct applicability of these equations to solve the full quantum problem is limited to the case of the quantum harmonic oscillator and quantum Brownian motion in a flat potential ($\mathcal{V}(\varphi)\propto \varphi$). Particularly, it cannot be used to describe coherent or incoherent quantum tunneling phenomena in anharmonic potentials. 
Nevertheless, the Eqs.~(\ref{Eq.langevinphs}) and (\ref{Eq.eqofmo}) are useful when they can be understood from a quasiclassical perspective, where quantum fluctuations only weakly alter the deterministic trajectory of a particle. In this quasiclassical approach, the operator-valued quantities are replaced by c-numbers but the fluctuating currents retain quantum statistics.

Within this context, the form (\ref{Eq.eqofmo}) makes the impact of the resistor $R_g$ on a system with total capacitance $C_{tot}$ and in a potential $\mathcal{V}(\varphi)$ easily accessible \footnote{Please note that we keep $C_{tot}$ fixed when we change the ratio $C/C_J$. In this way, changing $C$ only changes the coupling to the anomalous environment and does not renormalize the intrinsic system properties.}.
We immediately deduce that for a constant current $I_b$, where $\mathcal{V}(\hat{\varphi}(t)) = -\Phi_0 I_b\, \hat\varphi(t) $  the time retarded damping part in Eq.~(\ref{Eq.eqofmo}) vanishes, while for a harmonic oscillator potential (i.e. a LC-circuit) we directly obtain a linear damping term 
yielding damped oscillations for the quasiclassical dynamics of $\varphi$ exactly as in the case of coupling to the phase, i.e. Eq.~(\ref{Eq.langevinphs}).

However, independently of the system under study, the fluctuating current $\hat{\boldsymbol{I}}_{Q}(t)$ is different to $\hat{\boldsymbol{I}}_{\varphi}(t)$ in Eq.~(\ref{Eq.langevinphs}). Since the thermal and quantum fluctuations of the  resistors alter the metastable decay we discuss in Sec.~\ref{Sec.meta}, this observation is of particular interest for this work. 
$\hat{\boldsymbol{I}}_{Q}(t)$ originates from the noise in the resistor $R_g$ but is enriched by the capacitive coupling. To see this we use that the fluctuating current affecting $\hat{\varphi}_*$ (see Fig.~\ref{Fig.lines}) is described by $\boldsymbol{\hat{I}}_{\varphi_*}(t)$ with the usual correlation function of the form 
\begin{align}
\langle \boldsymbol{\hat{I}}_{\varphi_*}(t)&\boldsymbol{\hat{I}}_{\varphi_*}(t')\rangle =  \frac{\hbar}{\pi}\int_0^\infty d\omega\frac{\omega}{R_{g}} f_C(\omega)\\ \nonumber&\times\left[\cos(\omega(t-t'))\coth\left(\frac{\beta\omega}{2}\right)-i\sin(\omega(t-t'))\right],
\end{align}
where we can identify the Ohmic spectral density with Drude cutoff function $f_C(\omega)=\left(1+\omega^2/\omega_R^2\right)^{-1}$ used for $R_g$ \footnote{ Note that  $\boldsymbol{\hat{I}}_{\varphi_*}$ coincides with $\boldsymbol{\hat{I}}_{\varphi}$ in Eq.~(\ref{Eq.langevinphs}) by replacing $R_g$ with $R_S$.}. The capacitive coupling translates the current $\boldsymbol{\hat{I}}_{\varphi_*}$ into 
\begin{align}
\boldsymbol{\hat{I}}_Q(t) \approx \frac{CC_{J}}{C_{tot}}\int_{0}^{t}du\,\eta(t-u)\boldsymbol{\dot{\hat{I}}}_{\varphi_*}(u),
\label{Eq.qforce}
\end{align}
where we neglected additional initial slippage terms by assuming $t\gg \tilde{\omega}_c^{-1}$.
Using Eq.~(\ref{Eq.qforce}), we find the correlation function
\begin{align}
    \langle \boldsymbol{\hat{I}}_{Q}(t)&\boldsymbol{\hat{I}}_{Q}(t')\rangle \approx \frac{\hbar}{\pi}\int_{0}^{\infty}d\omega \frac{C^{2}R_{g}\omega^{3}}{(1+\frac{\omega^{2}}{\tilde{\omega}_{c}^{2}})}f_C(\omega)\\ \nonumber&\times\left[\cos(\omega(t-t'))\coth\left(\frac{\beta\omega}{2}\right)-i\sin(\omega(t-t'))\right],
\end{align}
where the factor in front of the bracket shows a non-linear behaviour in $\omega$ stemming from the translation from phase-noise acting on $\varphi_*$ to charge-noise affecting $\varphi$. Please note that this equation has different high frequency regimes.  For $\omega_R \gg\omega\gg\tilde\omega_c$, we find that the capacitor $C$ becomes transparent leading to an Ohmic behaviour for the correlation function
\begin{align}
    \frac{C^{2}R_{g}\omega^{3}}{(1+\frac{\omega^{2}}{\tilde{\omega}_{c}^{2}})}f_
    C(\omega)\approx  \frac{C^{2}R_{g} }{\tilde\omega^2_c} \omega  f_C(\omega)
\end{align}
while for $\omega\gg \omega_R\gg\tilde\omega_c$ we obtain furthermore $\omega f_C(\omega)\rightarrow 0$ i.e. the resistor $R_g$ has no influence in this frequency regime. We will be particularly interested in the regime $C\gg C_J$ which yields in the limit $\omega_R\rightarrow\infty$ 
\begin{align}
   \frac{C^{2}R_{g}\omega^{3}}{(1+\frac{\omega^{2}}{\tilde{\omega}_{c}^{2}})}f_C(\omega)\approx \frac{C^2R_g\omega^3}{1+C_J^2R_g^2\omega^2},
\end{align}
where $C_J$ leads to a natural high frequency cutoff for the the super-Ohmic behaviour. 

A more rigorous derivation of the quasiclassical Langevin equations can be obtained from a semiclassical expansion within the full Feynman-Vernon path integral description \cite{Weiss2012}.
%
%
%
%
%
%
%
\begin{figure*}[t]
	\centering	\includegraphics[scale=0.071]{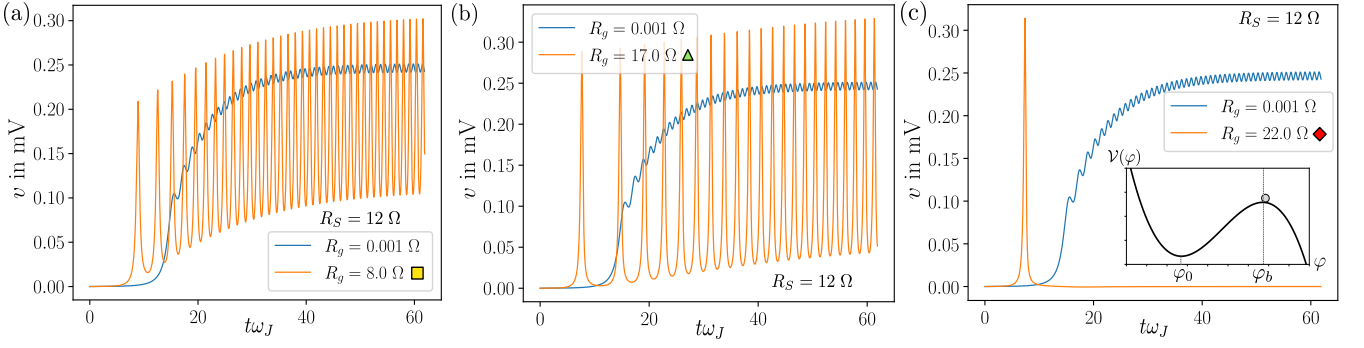}
	\caption{Voltage dynamics of the current biased Josephson Junction as a function of time. The particle starts at rest close to a maximum in the tilted washboard potential as we display in the  inset of (c). We fix $i_b=0.98$ and $R_S=12\;\Omega$ ($\gamma/\omega_J = 0.13$). The blue lines in each plot are guides to the eye and show the results for vanishing $R_g$. (a) $R_g=8 \, \Omega$ ($\tau_p\omega_J = 4.9$) and (b) $R_g=17\, \Omega$ ($\tau_p\omega_J = 10.41$): The oscillations in the voltage stemming from the washboard are strongly amplified by $R_g$. (c) $R_g = 22\; \Omega$ ($\tau_p\omega_J = 13.47$): The particle is first accelerated and then re-trapped in the next local minimum of the potential. This happens only due to the combined effects of $R_g$ and $R_S$. The symbols in the respective legends relate the shown parameters to Fig.~\ref{Fig.colourstate} (see below).  Circuit parameters: $C/C_J = 0.01$, $\, I_C= 21\;\mu$A, $C_{tot}=6$ pF, $I_b/I_C = 0.98$.   }
	\label{Fig.voltage}
\end{figure*}

\subsection{Dynamics of the generalized RCSJ model: Voltage state and re-trapping}

We now proceed by analyzing the phase dynamics in the presence of both dissipative couplings in the semiclassical approximation discussed in the previous section.
%
As a paradigmatic example, we study the  dynamics of the Josephson phase within a washboard potential. Hence,  we choose  $\mathcal{V}(\varphi) = - \Phi_0  I_b \varphi  - \Phi_0 I_C \cos(\varphi)$, where $I_b$ is the current bias and $I_C$ the critical current of the JJ (see also \cite{Martinis1990}). In this potential, classical and quantum  fluctuations are important for escaping local minima by thermal hopping or quantum tunneling and for re-trapping a running state in another minimum (phase diffusive dynamics) \cite{Vion1996}.

Technically, a stepwise strategy is used here and in the sequel to describe the full dynamics: We first calculate the deterministic dynamics in the washboard potential and then discuss the impact of fluctuations separately. We review this procedure in detail in the next subsection (see Sec.~\ref{Sec.fluctuations}). 
First, the relevant frequencies and timescales are considered, i.e. 
\begin{align}
    \tau_p &= R_gC \nonumber\\
    \gamma & = 1/(R_SC_{tot}) \\
    \omega_J &= \sqrt{I_C/(\Phi_0 C_{tot})}\nonumber\, , 
\end{align}
in order to make  time   $s= t\,\omega_J$, currents $i_b=I_b/I_C$, and  capacitances $c_0 = C/C_{tot}$ dimensionless. Furthermore, we assume both resistors to have strictly Ohmic spectral densities via $\omega_R/\omega_J\rightarrow \infty$. 
Now, for the deterministic trajectories, expectation values with respect to the quantum noise of the combined semiclassical Langevin equation are taken with an equilibrium initial state for the bath and the system. 
In this way, the initial slippage terms vanish together with the fluctuating currents and one arrives at
\footnote{Hence, for pure charge dissipation this means that we have $\langle\hat{I}_0\rangle = - \langle\hat{I}_g\rangle$ i.e. that the current through the $C$ -- $R_g$ branch is exactly opposite to the one through the junction.}
\begin{align}
    \ddot{\varphi}(s)=& i_b-\sin(\varphi(s))-\frac{\gamma}{\omega_J}\dot{\varphi}(s)  - c_0 \gamma  \int_0
^s \!\!d\nu\, \eta(s-\nu) \ddot{\varphi}(\nu)\nonumber \\
    &\!\!- c_0 \int_0
^s \!\!d\nu\, \eta(s-\nu) \cos(\varphi(\nu)) \dot{\varphi}(\nu).
\label{eq:mostgen} 
\end{align}
This equation still includes the retardation effects originating from $\tilde{\omega}_c/\omega_J \approx c_Jc_0(\tau_p\omega_J)^{-1} $, where  $c_J = C/C_J$. 
By  taking the limit $C\gg C_J$ ($c_0 \rightarrow 1,c_J \gg 1 $), we find a regime where $\tilde{\omega}_c/\omega_J\gg 1$ and therefore time-local response $\theta(s)\eta(s) \approx \tau_p \omega_J \delta(s)$, where $\theta(s)$ is the Heavisde step function. This leads to the simplified equation
\begin{align}
\left(1+\gamma \tau_p\right)\ddot{\varphi}(s) & \approx {i_{b}}-\sin(\varphi(s))\nonumber\\&-\!\left(\frac{\gamma}{\omega_J} +\tau_p\omega_J\cos(\varphi(s))\right)\dot{\varphi}(s).\label{Eq.Langevinfull}
\end{align}
Physically, in this limit the charge $Q$ falls off almost completely at the the capacitor $C$, making the impact of $R_g$ maximal because the system cannot avoid the slowing down of charge relaxation defined by $\tau_p$. 

Typically, the effect of $R_g$ is explained as \textit{frequency dependent damping} which can be intuitively deduced by arguing that the capacitor $C$ blocks low frequency contributions \cite{Martinis1990}. However, we refine this statement for the case of a current biased JJ by carefully looking at Eq.~(\ref{Eq.Langevinfull}). Because, the frequency changes sign along the washboard potential, we find that the effect of the resistor $R_g$ depends on the phase difference via $\cos(\varphi(t))$, effectively realizing \textit{damping or amplification} in this limit.
While damping appears for $\cos(\varphi(t)) > 0$, the fictitious particle accelerates for $ \cos(\varphi(t)) < 0$. 
%
%
%
%
%
%
\begin{figure*}[t]
	\centering
    \includegraphics[scale=0.2835]{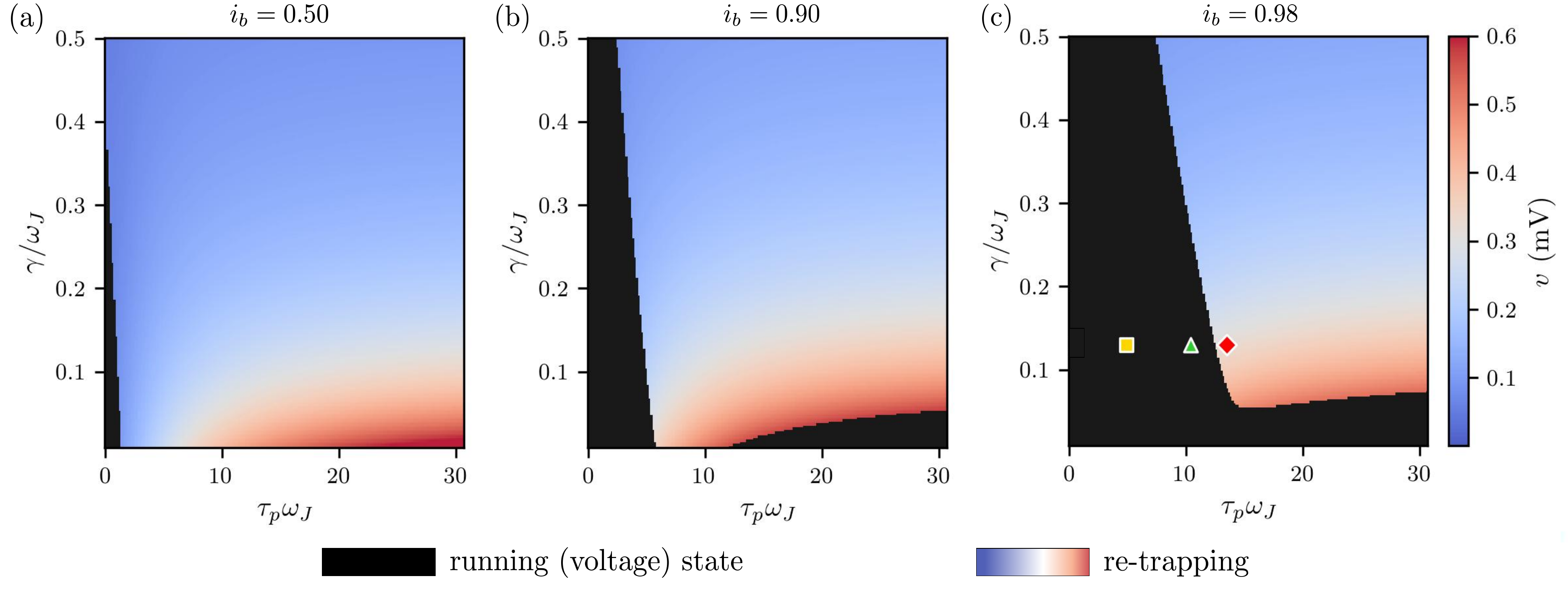}
	\caption{Colour-plots indicating the deterministic running  state (black areas) with finite voltage and the deterministic  re-trapping regime (coloured areas) of the junction as a function of the dissipative parameters $\gamma/\omega_J$ and $\tau_p\omega_J$ for different values of the bias current. The colour code encodes the height of the voltage pulse across the junction before re-trapping (see Fig.~\ref{Fig.voltage}c). (a) $i_b = 0.50$: The particle is re-trapped for $\gamma/\omega_J\rightarrow 0$ and finite $\tau_p\omega_J$ and vice versa. Having both environments present, the system is re-trapped for a broad parameter regime. The height of the voltage pulse before re-trapping increases with $\tau_p\omega_J$. (b) $i_b= 0.9$: In the depicted parameter regime phase dissipation alone does not lead to re-trapping. For low $\gamma/\omega_J$, we find a transition to trapping and then a transition back to the running state by increasing $\tau_p \omega_J$. (c) $i_b= 0.98$: Strongly tilted washboard potential. Both resistors are needed to achieve trapping in the next potential minimum. The symbols show the dissipative coupling of the examples displayed in Fig.~\ref{Fig.voltage}.    Further parameters: $I_C=21\,\mu$A, $C_{tot}=6$ pF, $C/C_J = 0.01$.}
	\label{Fig.colourstate}
\end{figure*}

We simulate the general equation of motion Eq.~(\ref{eq:mostgen}), which includes the retardation effects and show the results for the voltage $v(t)=\frac{\hbar}{2e}\dot{{\varphi}}(t)$  for large bias $i_B=0.98$ in Fig.~\ref{Fig.voltage}. There, we fix $\omega_J$ by choosing the system parameters  $I_C=21\,\mu$A, $C_{tot}=6$ pF, and furthermore set $C/C_J = 0.01$ and $R_S=12\,\Omega$ ($\gamma/\omega_J=0.13$). 
As initial conditions, we set $\varphi(0) = \varphi_b+\epsilon$ and $\dot{\varphi}_0=0$, where $\varphi_b$ is the top of the barrier as defined in Fig.~\ref{Fig.voltage}c and $\epsilon$ is a small positive number that ensures that the particle begins to move in the potential. 
We  compare the time evolution  of $v(t)$ for $R_g = 8$ ($\tau_p\omega_J = 4.9$),  $R_g = 17\,\Omega $ ($\tau_p\omega_J = 10.41$) and $R_g = 22 \,\Omega $ ($\tau_p\omega_J = 13.47$) in Figs.~\ref{Fig.voltage}a-c, respectively. Please note that for the parameters chosen here (i.e. $I_C$ and $C_{tot}$) already a low Ohmic environment leads to strong coupling to the charge environment as indicated by the large values of $\tau_p \omega_J$.

In Figs.~\ref{Fig.voltage}, we see that the non-linear damping behaviour amplifies the oscillations in the voltage produced by the movement of the phase particle in the washboard potential. Interestingly, a regime can be identified where the phase particle starts to run down the washboard, is amplified by charge dissipation and then localizes in the next minimum (see Fig.~\ref{Fig.voltage}c).
A particle that escapes from a local metastable minimum by thermal hopping or macroscopic quantum tunneling (MQT) (see next section), then accelerates and gets re-trapped in the next metastable well. This process repeats so that the dynamics consists of sequences of  hopping (phase slips) and re-trapping events, called phase diffusion. We comment on the corresponding parameter regime for this  dynamics in the next subsection.
Note, that for realistic  parameters, this scenario happens even in the regime $I_b\sim I_C$. 
Typically, in this regime and for $R_g=0$, re-trapping can only be described by taking the thermal or quantum fluctuations in the Langevin equation into account \cite{MELNIKOV1991,Belzig2015}.\\

Thus, in order to obtain a more complete picture, we further discuss the re-trapping behaviour of the system in the entire parameter regime. We again assume the phase particle at $t=0$ to be at rest close to the top of a barrier in the washboard potential and  calculate the time evolution for different bias currents as a function of the dissipative couplings. We detect the parameter-regime where the particle is re-trapped in the next metastable well and show results in Fig.~\ref{Fig.colourstate}. There the regime of the running (voltage) state is shown in black and the regime of re-trapping is indicated by the coloured region. Furthermore, the colour bar indicates the height of the voltage pulse that occurs just before the re-trapping (see Fig.~\ref{Fig.voltage}c).

Interestingly, we observe a non-monotonic re-trapping behavior in Figs.~\ref{Fig.colourstate}b and c.
For low $\gamma/\omega_J$, we find that by increasing $\tau_p\omega_J$, the system changes from a running state to trapping and back to a running state.
This can be explained by retardation effects originating from $\tilde{\omega}_c/\omega_J \sim 1$ for large $\tau_p\omega_J$. It implies that for large $\tau_p\omega_J$ the damping/acceleration as described by  Eq.~(\ref{Eq.Langevinfull}) is not a good approximation and the anomalous environment cannot follow the dynamics fast enough to lead to localization. However, by increasing $\gamma/\omega_J$, the phase dynamics  slows down and the particle becomes again trapped in the next minium.
The symbols in Fig.~\ref{Fig.colourstate}c connect to the parameters of the simulation shown in Figs.~\ref{Fig.voltage}a-c.  

\subsection{Impact of fluctuations on the dynamics} \label{Sec.fluctuations}
After the discussion of the deterministic part of the evolution, we now include the impact of quantum or thermal fluctuations.  These fluctuations predominantly matter close to the barriers, where thermal hopping or quantum tunneling is important for the dynamics.  

%
%
%
%
%
%
\begin{figure*}[t]
	\centering
    \includegraphics[scale=0.0376]{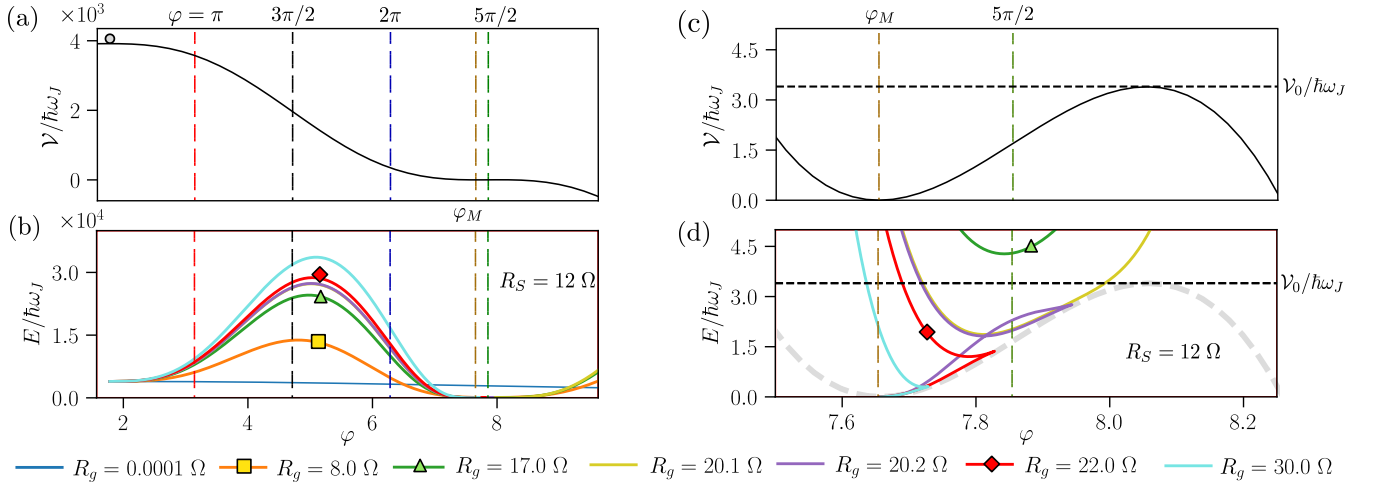}
	\caption{(a) Exact shape of the tilted washboard potential for $i_b=0.98$. The particle starts close to the barrier top. The dashed lines indicate the impact of the anomalous dissipation as determined by Eq.~(\ref{Eq.coupling}):  $\varphi=\pi$ maximum amplification,  $\varphi=3\pi/2$ sign change, $\varphi=2\pi$ maximum damping, $\varphi=5\pi/2$ sign change. The golden dashed line shows the  potential minimum ($\varphi_M$) adjacent to the barrier. 
    (b) Total instantaneous energy of the particle at position $\varphi$ for different values of $R_g$ (different lines) and $R_S=12\, \Omega$. The dashed lines are guides to the eye (see (a)). (c) Zoom into the region of the first minimum with the adjacent barrier. 
    (d) Total instantaneous energy in the barrier region for different values of $R_g$ and $R_S= 12\,\Omega$. While for $R_g=20.2 \; \Omega$ the particle falls back into the minimum for $R_g=20.1 \; \Omega$ the particle has enough energy left to transition over the barrier due to the amplification originating from the anomalous coupling effectively lowering the activation energy. The gray dashed line shows the potential energy alone. The symbols in (b) and (d) connect to  Figs.~\ref{Fig.voltage} and \ref{Fig.colourstate}.  }
	\label{Fig.energystate}
\end{figure*}
We deal with the escape rates out of a metastable minimum originating from quantum and classical fluctuations in detail in the next section.
However, there we assume the phase to be initially trapped in one metastable well and to be in equilibrium with the environment.
While this is perfectly justified for the  escape of an initially prepared particle the impact of the fluctuations on the subsequent dynamics and specifically the re-trapping behaviour after the escape process is generally more complicated.
Especially, phase diffusive regimes, where the particle dynamics is described by statistical hopping between minima gained a lot of attention over the last years \cite{Haxell2023}.   

To obtain a qualitative understanding of the impact of fluctuations on the re-trapping, we provide a phenomenological discussion in this section. Thereby, we consider only temperature regimes $k_BT\ll \mathcal{V}_0$ where $\mathcal{V}_0$ is the barrier height adjacent to a local minimum (see Fig.~\ref{Fig.energystate}c) and the limit of strong bias $i_b = 0.98$, where deterministic re-trapping with pure phase dissipation is hard to achieve as we showed in Fig.~\ref{Fig.colourstate}c.

To discuss the statistical re-trapping  for different damping regimes, we use  Eq.~(\ref{Eq.Langevinfull}) (valid in the regime $\tilde{\omega}_C/\omega_J\gg1$) to define the total damping parameter  
\begin{align}
    \alpha({\varphi}(t)) = \frac{\left(\frac{\gamma}{\omega_{J}}+\tau_{p}\omega_{J} \cos(\varphi(t))\right)}{\left(1+\gamma\tau_{p})\right)}\label{Eq.coupling}.
\end{align}
For the results displayed in  Fig.~\ref{Fig.colourstate}, the particle is  underdamped for $\alpha <1$ corresponding to the running state regime for small $\tau_p\omega_J$. There, it is well known that the fluctuations can lead to re-trapping only with an exponentially small rate $\Gamma_{\text{rt}}$ followed by a subsequent escape process with rate $\Gamma$ as defined in Sec.~\ref{Sec.3} \cite{MELNIKOV1991}.\\ \indent
Because the impact of charge dissipation depends on $\cos({\varphi})$, there is no overdamped regime ($\alpha \gg 1$) for a full trajectory  (assuming $\gamma/\omega_J <1$ as in Fig.~\ref{Fig.colourstate}). However, to understand whether the fluctuations can significantly alter the deterministic re-trapping behaviour described by the coloured region in Fig.~~\ref{Fig.colourstate}c, we  look at the physics close to the minimum before the barrier. 
Inserting a minimum of the washboard potential $\varphi_M=\arcsin(i_b)+2\pi \,n$ (with $n\in \mathbb{N}$) into the total damping parameter, we obtain $\cos(\varphi_M)\approx 0.2$ for $i_b = 0.98$ suppressing the impact of $\tau_p \omega_J$. 

Nevertheless, for sufficiently large $\tau_p \omega_J$, there is a regime  $\alpha(\varphi_M) \gg 1$, where the physics is described by fast equilibration in the first well irrespective of the initial conditions. In this limit, the inertia term in the Langevin equation can be neglected, and the system cannot climb any barrier. Hence, the fluctuations only lead to an escape rate from an equilibrium state. This regime corresponds to the upper right corner in Fig.~\ref{Fig.colourstate}c \footnote{Note that Eq.~(\ref{Eq.Langevinfull}) is not valid in the lower corner of Fig.~\ref{Fig.colourstate}c where retardation effects become important}. 

However, even a moderately damped particle at position $\varphi_M$  (i.e. $\alpha(\varphi_M) \sim 1$) can behave as being overdamped. 
In this parameter regime looking  only at the position $\varphi_M$ is misleading, because the regime with maximal damping (i.e. $\cos(\varphi)=1$, displayed by the blue dashed line in Fig.~\ref{Fig.energystate}a) comes before the minimum. Hence, the particle may come from an overdamped regime having already lost a significant amount of energy approaching the minimum. 

On the other hand, the regime where the cosine changes sign is very close to the minimum (compare golden and green dashed lines in Fig.~\ref{Fig.energystate}a).
So, even if the velocity close to the valley is comparatively low, fluctuations might bring the particle into a regime where it is no longer damped but even accelerated by the dissipative coupling. These circumstances make the straightforward assessment of the impact of the fluctuations on the dynamics in the regime where $ \alpha(\varphi_M) \sim 1 $ difficult.

To obtain a better understanding of how fluctuations can alter the dynamics, we plot the instantaneous (classical) energy of the phase particle $E=E_{kin}(\dot{{\varphi}}) + \mathcal{V}({\varphi})$ as a function of $\varphi$ for different values of $R_g$ in Fig.~\ref{Fig.energystate}b and d, using dissipative coupling parameters around the values discussed in Fig.~\ref{Fig.voltage}. Discussing $E$, we can deduce the activation energy that must be provided by the fluctuations for the particle to cross the barrier.
%
%
%
%
%
%
%
%
\begin{figure}[t]
	\centering
    \includegraphics[scale=0.89]{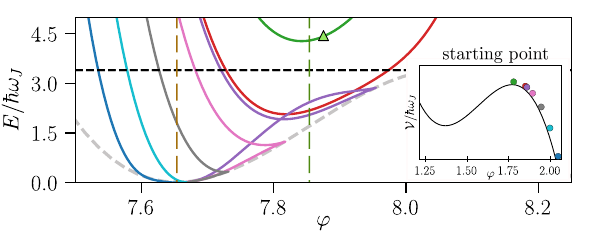}
	\caption{Total energy of the phase particle close to the first minimum of the washboard potential (compare to Fig.~\ref{Fig.energystate}c and d) for $R_g=17\; \Omega$ and    $R_S=12\; \Omega$ for different initial values $\varphi_0$ (different lines). The respective initial values are shown in the inset. The initial value on top of the barrier corresponds to the situation labelled by the green triangle and connects to Figs.~\ref{Fig.voltage}, \ref{Fig.colourstate} and \ref{Fig.energystate} }
	\label{Fig.17example}
\end{figure}

By comparing the energy in Fig.~\ref{Fig.energystate}b with the respective position in the washboard potential shown in Fig.~\ref{Fig.energystate}a, we see how the charge dissipative environment changes the total energy in the system along the trajectory.
While we show a zoom into the first valley and barrier region of the washboard potential in Fig.~\ref{Fig.energystate}c, we display the corresponding total energies in  Fig.~\ref{Fig.energystate}d. We find that all trajectories surpass the minimum. Thereby, all values below $R_g=20.2\,\Omega$ are not re-trapped but also overcome the barrier. The green dashed line shows where the impact of charge dissipation changes sign. 
Interestingly, the value $R_g=20.1 \; \Omega$ has enough energy left to enter a regime where the charge coupling pushes the particle over the barrier. Changing $R_g$ only slightly (to $R_g=20.2 \; \Omega$), the acceleration originating from charge dissipation is not enough to overcome the barrier, and the particle falls back into the minimum. Please note that here (violet line), the equilibration to the minimum is accelerated, as can be seen from the increasing total energy of the trajectory going back to $\varphi_M$. 
Further increasing $R_g$, we approach the overdamped regime, as indicated by the distance of the turning points of the trajectories to $\varphi_M$  for $R_g>20.2\, \Omega$.
From these results, we see that the sign change of the charge dissipation effectively makes the activation energy for overcoming the barrier position dependent, making the impact of the fluctuations more prominent.

In principle, for each circuit parameter, the new activation energy has to be calculated independently. As an example, we show in Fig.~\ref{Fig.17example} the total energies for $R_g=17\,\Omega$ for different initial values $\varphi_0$.
Depending on where the particle starts, the kinetic energy in the first minimum is different. The deterministic equation of motion shows that starting at the barrier top (green line) yields a running state, while starting further below, the particle is re-trapped.
To define a position dependent activation energy for particles that are re-trapped, the energy of their turning point (i.e. the point where we have $E_{kin} = 0$) has to be compared with a trajectory with minimal $E_{kin} \neq 0$ that is still traversing the barrier.
In Fig.~\ref{Fig.17example}, an example for a small position dependent activation energy is found by the energy difference of the red and the violet line at the turning point of the latter. In this regime, fluctuations play an important role for the dynamics, as only small fluctuations change the final state of the system.
However, for the trajectories that end close to the minimum the activation energy is still far away and fluctuations do not have a significant impact as long as $k_BT\ll \mathcal{V}_0$ yielding a trapping probability $P_\text{tr}\sim1$.
Note that the deterministic results in Fig.~\ref{Fig.colourstate} have to be understood in the context discussed in this section, which means that  the boundary between the deterministic running solution and the trapping solution is strongly affected by fluctuations.  

We see this also from  the results in  Fig.~\ref{Fig.17example}, which indicate a strong dependence of the dynamics on where the particle escapes the initial metastable minimum. 
While initial values at the top emulate thermal hopping, initial values below this value emulate quantum tunneling. \\
In the following, we give a detailed analysis how a previously trapped particle escapes a metastable minimum and particularly also study the most probable escape points beyond the barrier. In this way, we obtain the full dynamics of the problem for  parameters that yield fast equilibration into the first minimum after escape. Hence, we describe a new kind of the phase diffusive behaviour in the strongly biased regime that strongly depends on the charge dissipative coupling.    

Finally, we mention that the regime  where the deterministic trajectory is close to overcome the barrier the impact of fluctuations promise rich dynamics, where a detailed discussion would require implementing  the full quasiclassical Langevin equation stochastically.

\section{Decay rates for the generalized RCSJ model: Theoretical framework} \label{Sec.3}
We now turn to  a situation in which the phase is initially trapped in one of the minima of the washboard potential. Particularly, we discuss a situation in which the local potential surrounding the phase can be described in the cubic approximation 
\begin{align}
\mathcal{V} \left( \varphi \right) \simeq \Phi_{0}^{2}C_{tot}\left(\frac{1}{2}\omega_{I}^{2}{\varphi}^{2}-\frac{1}{3}\omega_{b}^{2}{\varphi}^{3}\right),
\end{align}
where $\omega_{I}\!=\!\omega_J\left(1-\left(I_{b}/I_{C}\right)^{2}\right)^{\frac{1}{4}}$
and $\omega_{b}\!=\!\sqrt{{I_{b}}/({2C_{tot}\Phi_{0}})}$.
This approximation is accurate when $I_{b}\rightarrow I_{C}$ and we will choose $I_{b}/I_{C}\geq0.98$ for all practical purposes throughout the following sections. Furthermore, we will work in the limit where $ \mathcal{V}_0 \ \gg \hbar \omega_I $, with $\mathcal{V}_0$ being the barrier height and $\omega_I$ the frequency of the metastable well. Please note that such a system is parametrized in the regime where the ratio between Josephson-
and charging energy is of the order of $E_J/E_C\sim 10^5$, with $E_C=4e^2/ C_{tot}$ and $E_{J}=\Phi_0 I_C$. Furthermore, we also stay in  the temperature regime where $\mathcal{V}_0  \gg k_B T$.

\subsection{Effective action and extremal paths}
At this point, we have to recall the most convenient theoretical framework to calculate dissipative switching rates $\Gamma$ out of the metastable potential. This so-called Im$F$-approach \cite{HanggiReview1990}, relates  the rate $\Gamma$  to the imaginary part of the free energy via
$
    \Gamma = -\frac{2}{\hbar}\,c_0\,\text{Im}\left(F\right),
$
where $F$  is the free energy $F=-k_B T \ln (Z)$, and $c_0 = T_0/T$ when the decay of the system is dominated by thermally activated decay and $c_0=1$ when the decay is dominated by quantum tunneling \cite{langer_statistical_1969,Weiss2012}. Here $T_0$ is the so-called crossover temperature defined below. 

To determine $ \Gamma$ thus necessitates to calculate the partition function $Z$ of an unstable system, where the small imaginary part reflects the metastability. This is most conveniently done  
 via the Euclidean path integral 
\begin{align}
Z=\mathcal{N} \oint \mathrm{D}[\varphi(\tau)] e^{-\frac{1}{\hbar} S_{\mathrm{eff}}[\varphi(\tau)]} \label{Eq.partfunc}
\end{align}
with $\mathcal{N}$ being a normalization constant. The full dissipative action is defined via
\begin{align}
S_{\text {eff }} [\varphi(\tau)]/\Phi_0^2 & =\! \int_{-\beta/2}^{\beta/2} \mathrm{d} \tau\left(\frac{ C_{\text {tot }}}{2} \dot{\varphi}^{2}(\tau)+\mathcal{V}[\varphi(\tau)]\right)  \\ & +\frac{1}{2}\!\iint_{-\beta/2}^{\beta/2} d \tau d \tau^{\prime} F^{(\varphi)}\left(\tau \!-\!\tau^{\prime}\right) \varphi(\tau) \varphi\left(\tau^{\prime}\right)\nonumber \\ \nonumber& +\frac{1}{2}\!\iint_{-\beta/2}^{\beta/2} d \tau d \tau^{\prime} F^{(Q)}\left(\tau \!- \!\tau^{\prime}\right) \dot{\varphi}(\tau) \dot{\varphi}\left(\tau^{\prime}\right) ,\label{Eq.action}
\end{align}
where $\beta= \hbar/(k_BT)$. The dissipative imaginary time kernel coupling to the phase reads by (Matsubara frequencies $\omega_n=2\pi n/\beta$)
\begin{align}
F^{(\varphi)}(\tau)=\gamma \frac{C_{tot} }{  \beta}  \sum_{n=-\infty}^{\infty} \left|\omega_{n}\right|f_R(\omega_n) e^{i \omega_{n} \tau},
\end{align}
where we again use the coupling strength $\gamma=1/({C_{tot}R_S})$ and the drude high frequency cutoff function $f_R(\omega_n)=\left(1+|\omega_l|/\omega_R\right)^{-1}$.
Furthermore, we express the kernel coupling to the charge as
\begin{align}
    F^{(Q)}(\tau)=-\frac{C}{ \beta} \sum_{n=-\infty}^{\infty} \frac{\tau_{p} {f_R\left(|\omega_{n}|\right)|\omega_{n}|}}{1+\tau_{p} |\omega_{n}|f_R\left(|\omega_{n}|\right)} e^{i \omega_{n} \tau},
\end{align}
with $\tau_p=R_g C$. For more information on the origin of these kernels we refer to the Appendix of \cite{Maile2022}.
The extremal paths of the action then follow from
\begin{align}
0 & = \Phi_0^2C_{tot} \ddot{\varphi}(\tau)+ \Phi_0^2\int^{\beta/2}_{-\beta/2} d \tau^{\prime} F^{(Q)}\left(\tau-\tau^{\prime}\right) \ddot{\varphi}\left(\tau^{\prime}\right) \nonumber \\ & -  \mathcal{V}'(\varphi(\tau))- \Phi_0^2\int^{\beta/2} _{-\beta/2} {d} \tau^{\prime} F^{(\varphi)}\left(\tau-\tau^{\prime}\right) \varphi\left(\tau^{\prime}\right).\label{Eq.EOM}
\end{align}
Here, the charge dissipative kernel couples to the acceleration $\ddot{\varphi}(\tau)$, which, in turn, strongly depends on the gradient of the potential. In fact, to make a closer connection to the real time version Eq.~(\ref{Eq.eqofmo}), we can use the limits $C/C_{tot}\approx 1$ and $\gamma = 0$ to rewrite the equation for the extremal path for only charge dissipation into
\begin{align}
    0=\,& \Phi_0^2C_{tot}\ddot{\varphi}(t)-\mathcal{V}'(\varphi(\tau) ) \nonumber \\&- \int^{\beta/2}_{-\beta/2} {d} \tau^{\prime} F^{(\tau_p)}\left(\tau-\tau^{\prime}\right) \mathcal{V}'(\varphi(\tau') ).
\end{align}
Here, we find an Ohmic kernel originating from $R_g$
\begin{align}
    F^{(\tau_p)}(\tau)= \frac{\tau_p} {  \beta}  \sum_{n=-\infty}^{\infty} \left|\omega_{n}\right|f_R(\omega_n) e^{i \omega_{n} \tau}
\end{align}
coupling to the gradient of the potential. Hence, also for the problem of metastable decay, we expect that the impact of anomalous dissipation depends strongly on the potential under study (see also \cite{maile_exponential_2021,Maile2022}).

\subsection{Formal rate expressions}\label{sec:ThermallyActivatedDecaymain}
In the high temperature limit the imaginary time domain $\beta$ becomes small and the only possible (meta-) stable solutions for the extremal paths in the partition function Eq.~(\ref{Eq.partfunc})
come from the vicinity of the stationary points of the action, i.e. of the time independent trajectory close to the well of the potential at $\varphi(\tau)=0$ and around the barrier top $\varphi(\tau)=\varphi_b$. Close to these points, we can perform a harmonic approximation on the action in Eq.~(\ref{Eq.action}) and correspondingly on the imaginary time equation of motion Eq.~(\ref{Eq.EOM}). We explain this in more detail in the Appendix \ref{sec:ThermallyActivatedDecay}. 
The stationary point $\varphi_b$ yields an imaginary part for the partition function Eq.~(\ref{Eq.partfunc}) leading to the escape rate \cite{langer_statistical_1969,Grabert1987}
\begin{align}\label{eq:EscapeRateAT0}
    \Gamma = f_q \, {\frac{\omega_I}{2\pi}\Theta_0 e^{-\frac{\mathcal{V}_0}{k_\text{B}T}}}\,\,
\end{align}
where we defined the scaled temperature $\Theta_0=2\pi k_\text{B} T_0/(\hbar\omega_I)$. The decay rate is formed by an classical exponential Arrhenius escape rate and the prefactor $f_q$ originating from the harmonic quantum fluctuations around the stationary points and defined in the Appendix~\ref{sec:ThermallyActivatedDecay}.

%
%
%
%
%
\begin{figure}[t]
	\centering
    \includegraphics[scale=0.26]{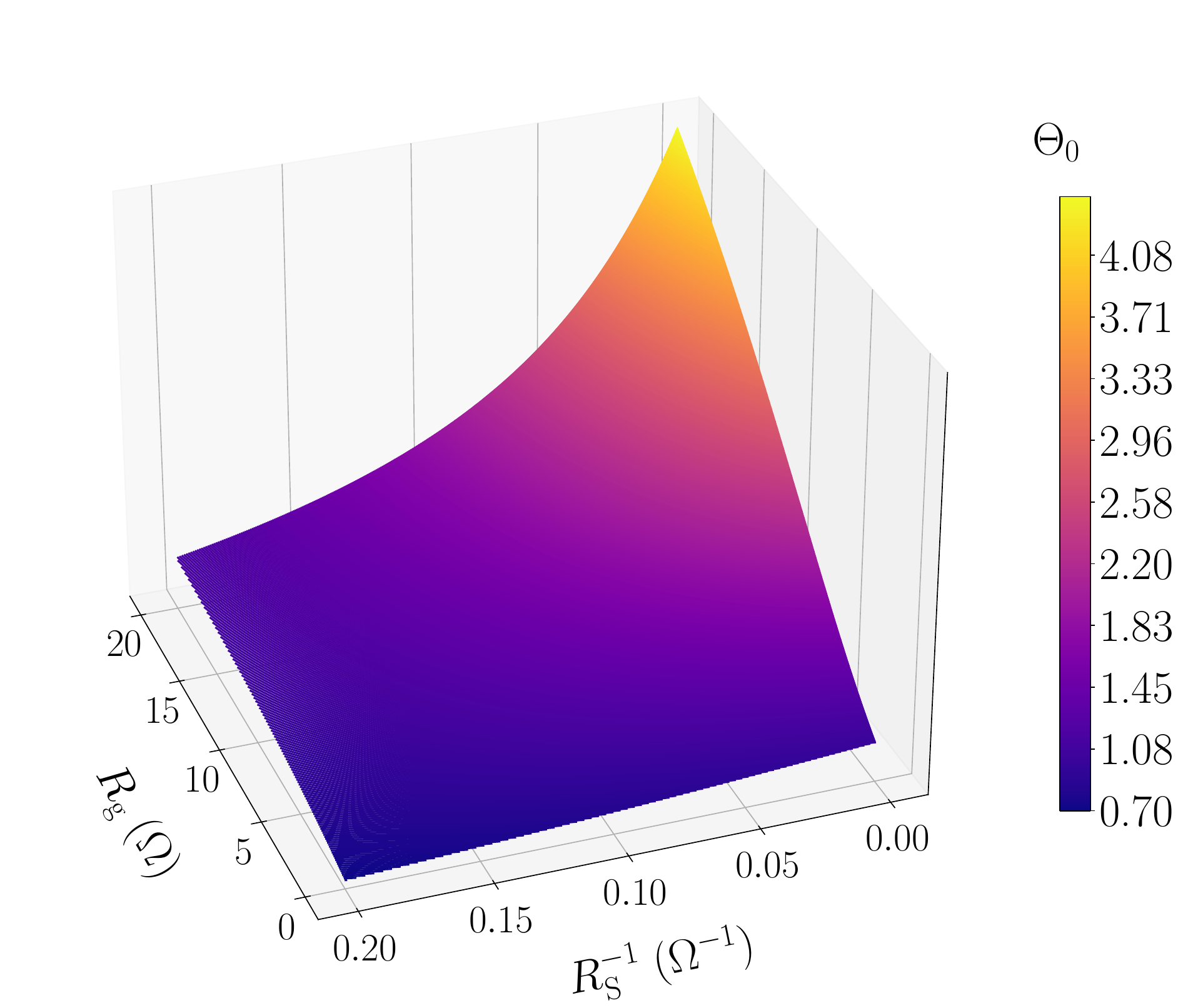}
	\caption{Crossover temperature as defined by Eq.~(\ref{Eq.crossover}) as a function of the resistances $R_g$ and $R_S$. While with increasing $1/R_S$ the crossover temperature  $\Theta_0$ decreases, with increasing $R_g$ it increases. Circuit parameters: $C/C_J = 0.01$, $\, I_C= 21\;\mu$A, $C_{tot}=6$ pF.}
	\label{Fig.crossovertemp}
\end{figure}

At low temperatures ($T \ll T_0$) the thermal activation over the barrier is exponentially suppressed and the particle is predominantly escaping via quantum tunneling \cite{Hanggi1984}. In this regime the result for the partition function again contains an imaginary part and may be written as  $Z=Z_0+iZ_B$. In the limit ($\mathcal{V}_0\gg \hbar \omega_I$) the information of the (most probable) barrier experienced by the tunneling particle is encoded in the most probable imaginary time path $\varphi_{B}(\tau)$. This path is defined via the extremum of the action $S_{B}=S_\text{eff}[\varphi_{B}(\tau)]$, dominating the imaginary part of the partition function
\begin{align}
    Z_B = \; \mathcal{F}_B\; e^{-S_B/\hbar},
\end{align}
where, the fluctuation prefactor $\mathcal{F}_B$ is defined in Eq.~(\ref{Eq.FB}) in the Appendix~\ref{Sec:Below}  and the path $\varphi_B(\tau)$ is defined via the full imaginary time equation of motion in Eq.~(\ref{Eq.EOM})
with conditions $\varphi_B(\beta/2)=\varphi_B(-\beta/2)\equiv\varphi_\mathcal{I}$ and the general symmetry $\varphi_B(\tau)=\varphi_B(-\tau)$.
We solve this equation numerically via  the technique introduced in \cite{grabert_temperature_1985,Grabert1987}. The solution is called \textit{bounce path}. As we can infer from Eq.~(\ref{Eq.EOM}), this path depends on both dissipative couplings and on the temperature. It turns out that the form of the path contains a lot of information we will leverage in the following sections. 
The rate for quantum tunneling is  readily calculated via 
\begin{align}
\Gamma = K \, e^{-S_{B}/\hbar},
\label{Eq.decay}
\end{align}
where $K$ is the so-called fluctuation prefactor we also  calculate numerically (see the Appendix~\ref{Sec:Below} and \cite{Grabert1987} for further information).

%
\section{Metastable decay: Explicit results} \label{Sec.meta}
In this section, we discuss the results obtained by applying the above methodology to metastable decay of the phase particle from a local minimum in the washboard potential. Thereby, we again fix $\omega_J$ by choosing $I_C=21\,\mu$A and $C_{tot}=6\,$pF. As explained above, in this regime, already values of the order of a few Ohm for $R_g$ lead to a significant impact on the system.

\subsection{The Crossover Temperature}
The breakdown of the harmonic approximation for the thermal decay is signaled by a divergence of the partition function at temperature $T_0$ due to a vanishing eigenvalue in the Gaussian fluctuation integral. Its value depends on the dissipative couplings and the circuit parameters, and is readily calculated to be
\begin{eqnarray}
\Theta_{0} &=&\frac{2\pi k_\text{B} T_0}{\hbar\omega_I}\nonumber\\
&=&\frac{\tau_{p} \omega_{I}\!-\!\frac{\gamma}{\omega_{I}}\!+\!\sqrt{\left(\frac{\gamma}{\omega_{I}}\!-\!\tau_{p} \omega_{I}\right)^{2}\!\!+4\left(1+\gamma \tau_{p}\right)}}{2\left(1+\gamma \tau_{p}\right)},\label{Eq.crossover}
\end{eqnarray}
where we used the limits $\omega_R/\omega_I\rightarrow \infty $ and $C_J/C\rightarrow 0$.
Although $T_0$ is not a temperature at which a physical divergence in the system can be observed, the breakdown of the technique occurs because the impact of the quantum fluctuations of the phase becomes too large to be captured by the harmonic approximation around the stationary points.
Hence, the change in the crossover temperature as a function of the dissipative couplings can be used to indicate the importance of quantum fluctuations in the system.  We display the results from Eq.~(\ref{Eq.crossover}) in the landscape plot in Fig.~\ref{Fig.crossovertemp} and find, as a first important result, that the resistor $R_g$ increases the crossover temperature. This means that the breakdown of the harmonic approximation happens at temperatures  $T_0(\tau_p\!\! \neq\! 0)>T_0( \tau_p\!\!=\! 0)$  indicating that quantum fluctuations play a larger role compared to the non-dissipative case.
Increasing $1/R_S$ the opposite happens. The quantum fluctuations of the phase are suppressed and the approximation is valid for lower values of $T$.
The effect of $R_g$ on $T_0$ has been already confirmed in experiments using superconducting interference devices for the readout of superconducting flux qubits \cite{Clarke2005} (see also 
\footnote{Please note that in their description the barrier frequency is defined with the bare $C_J$ and not as in our case with $C_{tot}$. Taking this into account both results are equivalent. Specifically, they show that reducing $R_g$ is reducing $T_0$, while we show that increasing $R_g$ is increasing $T_0$}).

%
%
%
%
\begin{figure}[t]
	\centering
    \includegraphics[scale=0.68]{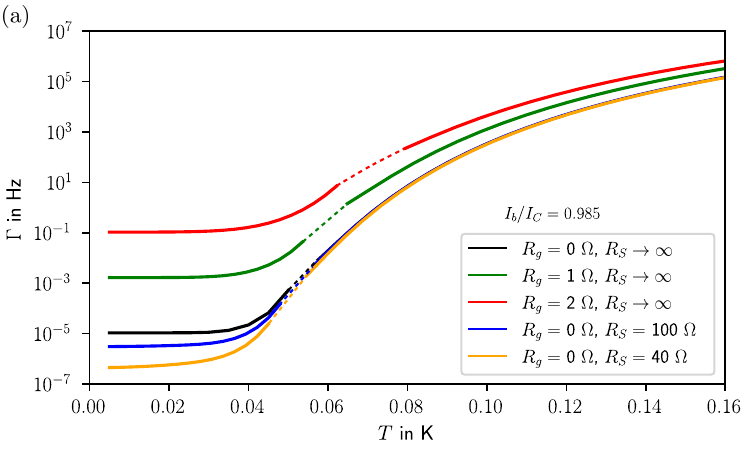}\\
    \includegraphics[scale=0.68]{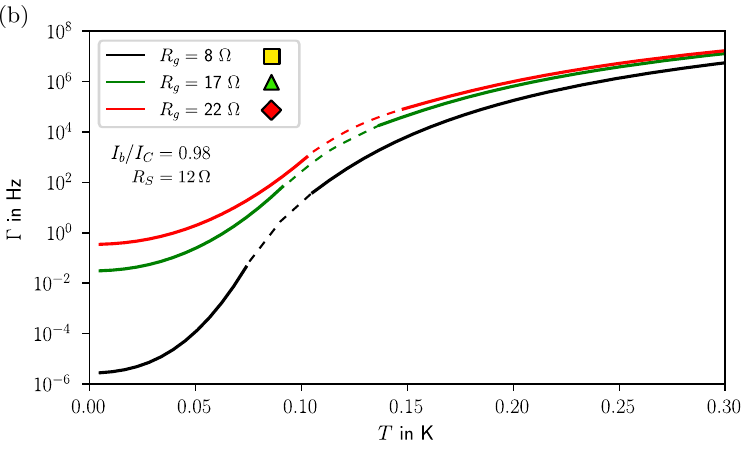}
	\caption{(a) Decay rate $\Gamma$ as a function of temperature for different values of the resistances $R_g$ or $R_S$. While phase (position) dissipation decreases the decay rate, charge (momentum) dissipation  increases it (here $
    C_J/C = 0.01$,  $I_b/I_C=0.985$). (b) Decay rate in the presence of both resistors $(C_J/C = 0.01$, $I_b/I_C=0.98)$. The symbols in the legend connect the respective parameters to Fig.~\ref{Fig.colourstate}c . }
	\label{Fig.Rate}
\end{figure}

\subsection{The decay rate as a function of temperature}
We use the introduced methods to calculate the decay rate above and below the crossover temperature, respectively, and show the results in the full temperature range in Fig.~\ref{Fig.Rate}. In Fig.~\ref{Fig.Rate}a, we study the case where either $R_g$ or $R_S$ is present. Furthermore, we show the non-dissipative case with the black line as a reference. The flat temperature-independent part at low temperature is the quantum tunneling regime, followed by thermal assisted tunneling indicated by a finite dependence of the rate on temperature. The dotted line is a guide to the eye interpolating the results for low temperatures to those for high temperatures. This regime is centred around the crossover temperature $T_0$, where neither the bounce technique nor the harmonic approximation around the stationary points is valid. 
The impact of phase dissipation is discussed regularly in the literature and leads to a suppression of the decay rate by decreasing $R_S$ (blue and orange line) \cite{Weiss2012,Golubev2019}. Furthermore, we can see the above discussed onset of temperature dependence of the rate by the finite slope of the line. For pure charge dissipation $R_g\neq0$ and $R_S\rightarrow \infty$ the opposite happens; the decay rate and the temperature regime of quantum tunneling is increased \cite{Weiss2012}. 
For all results we find an enhancement of the rate above the crossover temperature compared to the Arrhenius law, originating from the quantum correction factor $f_q$ in Eq.~(\ref{eq:EscapeRateAT0}). Please note that the dissipative couplings to the resistors change the decay rate by several orders of magnitude. We show this in more detail in  Figs.~\ref{Fig.EnResults}c and d, where we study the  quantum tunneling rate as a function of the dissipative couplings for different temperatures (see also the next section). 
%
%
%
\begin{figure}[t]
	\centering
	\includegraphics[scale=1.2]{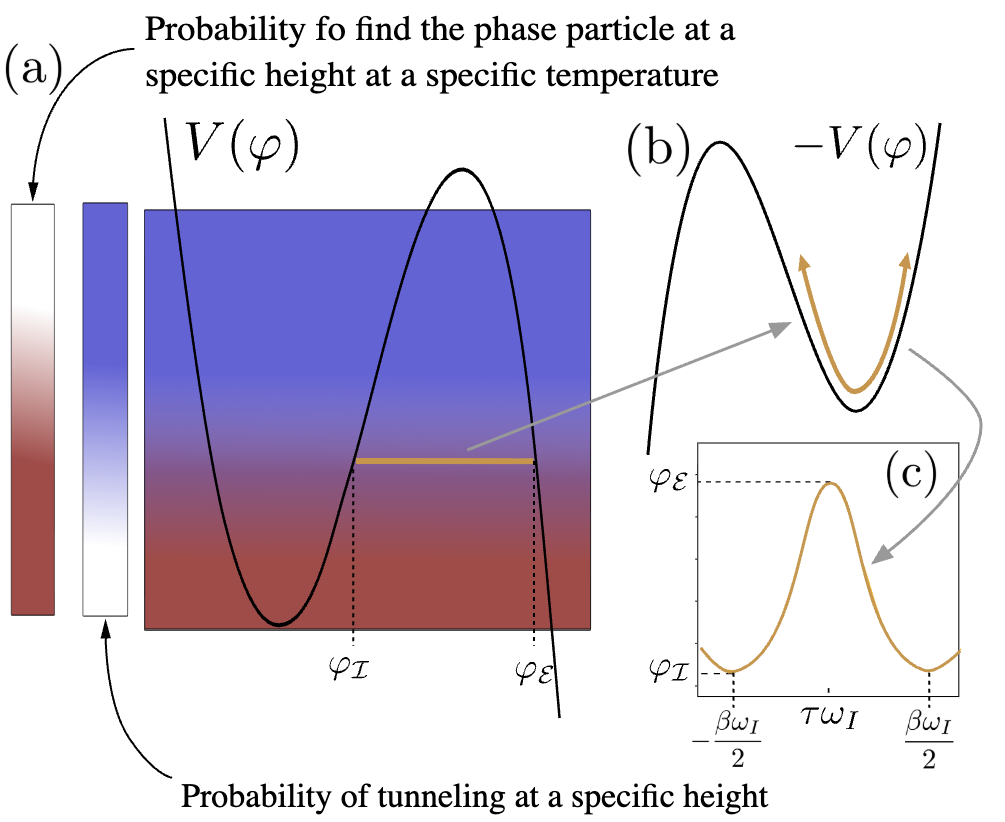}
	\caption{ Schematic illustration of the bounce path for finite temperatures. (a) The most probable energy  for barrier penetration originates from a trade-off between thermal and quantum fluctuations. While the thermal probability to find the particle at a specific energy decreases exponentially, the probability of tunneling increases exponentially for higher energies as the effective barrier becomes less opaque. The strongest overlap of both competing probabilities is indicated by the violet region. (c) The bounce corresponds to the periodic path $\varphi_B(\tau), \tau\in \{-{\beta}/{2}, {\beta}/{2}\}$ in the inverted potential (b) with boundary conditions according to $\varphi_\mathcal{I}\equiv\varphi_B(\pm\beta/2)$ and $\varphi_\mathcal{E}\equiv\varphi_B(0)$. 
	}
	\label{Fig.bouneexp}
\end{figure}
%
%
%
In presence of both dissipative couplings, a mixture of the two findings emerges. This is depicted in Fig.~\ref{Fig.Rate}b, where we discuss the same parameters as in Fig.~\ref{Fig.voltage}, as indicated by the symbols in the legend.  Particularly, we find that the rate is still enhanced by several orders of magnitude. In contrast to Fig.~\ref{Fig.Rate}a, a linear interpolation between the high and low temperature results is not a good approximation, indicating a more interesting crossover-regime, as in the case of pure phase dissipation. This might be connected to position dependent activation energy in the well region as discussed in Sec. \ref{Sec.2}. We leave the detailed study of this temperature regime for future work.

\begin{figure*}[t]
	\centering
	\includegraphics[scale=0.90]{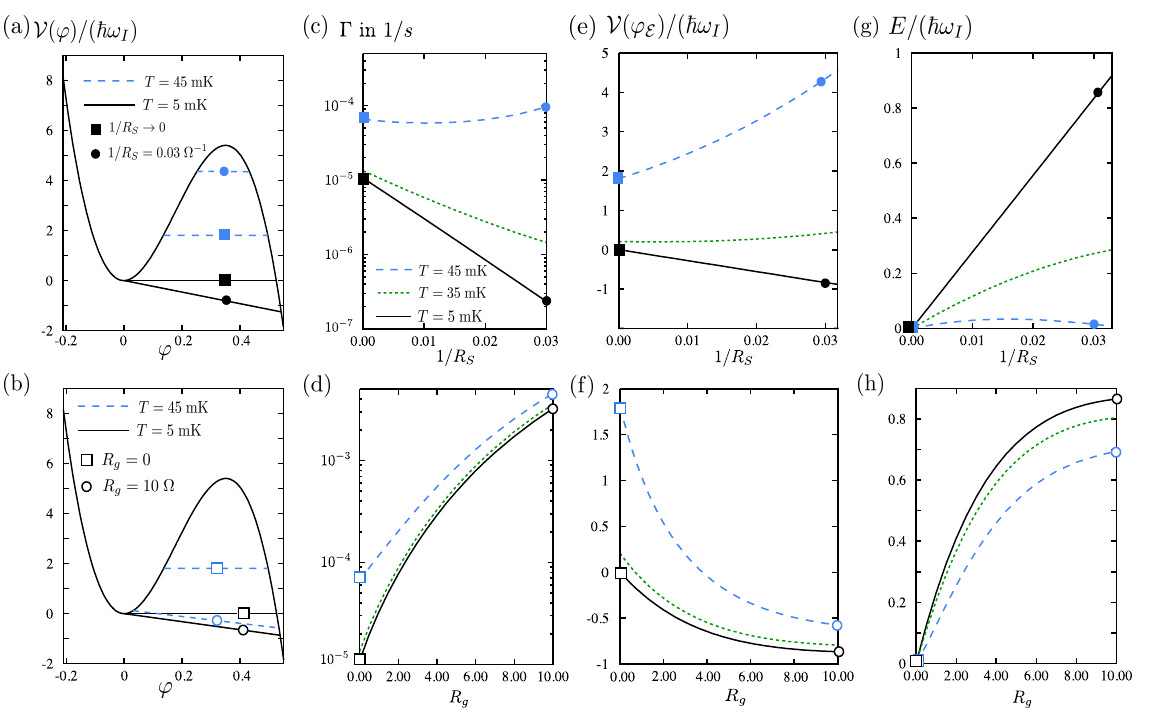}
	\caption{Effects of $R_S$ (top row) and $R_g$ (bottom row) on the tunneling process. Here, $C=C_J$, which suppresses the effect of $R_g$ compared to Fig.~\ref{Fig.Rate}, and $I_b/I_C = 0.985$. (a) and (b): Sketch of the cubic potential barrier. We display the most probable imaginary time trajectories by connecting the most probable entry point $\varphi_\mathcal I$ with the most probable escape point $\varphi_\mathcal{E}$. (a) Comparison of the dissipative trajectory for $1/R_S = 0.03 \,\Omega^{-1}$  (labelled by a circle) with the non-dissipative trajectory (labelled by a square) for two different temperatures $T=5$ mK (black solid line) and $T=45$ mK (blue dashed line). While for low temperatures the dissipative trajectory lies energetically below the non-dissipative one, for high temperatures the behaviour is opposite and the particle predominantly tunnels at higher energies due to the strong suppression of quantum fluctuations by phase dissipation.  (b) Analogous plot for charge dissipation with $R_g = 10$ $\Omega$. The dissipative particle always tunnels at lower energies compared to the non-dissipative one. This is consistent with the interpretation that $R_g$ enhances quantum fluctuations in the system. (c) and (d): Escape rate as a function of the dissipative couplings for three different temperatures. (e) and (f): Most probable escape points as a function of $1/R_S$ and $R_g$. (g) and (h): Most probable energy $E/(\hbar \omega_I)$ dissipated into the respective resistor as a function of $1/R_S$ and $R_g$. The symbolic labels for the trajectories in (a) and (b) connect to the accordingly labelled parameters in  (c) - (g). }
	\label{Fig.EnResults}
\end{figure*}

\subsection{The instanton bounce and its relation to the most probable escape path}

In the quantum tunneling regime $(T<0.9 \;T_0)$, the action of the bounce path $\varphi_B(\tau)$ is the dominant contribution to the imaginary part of the partition function Eq.~(\ref{Eq.partfunc}) and therefore determines the decay rate. In this section, we will take a closer look at the details of the path $\varphi_B(\tau)$. Its mathematical origin can be understood as the classical trajectory in the inverted potential $-\mathcal{V}(\varphi)$, with an initial value $\varphi_I(\beta)$ and the maximum value $\varphi_\mathcal{E}(\beta)$, that depend on $\beta$.
Generally, the bounce path is the  trajectory that probes the most important part of the barrier for the tunneling problem. In this context -- keeping in mind that it is a trajectory along imaginary time --  we can  use the bounce path to  understand where the particle is most likely to penetrate the barrier.   
In this context, we call the values $\varphi_\mathcal{I}$ and $\varphi_\mathcal{E}$ \textit{entry} and \textit{escape} points and define the most probable escape path by connecting $\varphi_\mathcal{I}$ with $\varphi_\mathcal{E}$, as displayed in Fig.~\ref{Fig.bouneexp}  (again, this is not to be understood as the real-time path the particle is taking).
The fact that in the non-dissipative case the path is vertical (i.e., $\mathcal{V}(\varphi_\mathcal{I})=\mathcal{V}(\varphi_\mathcal{E})$) is readily interpreted as the tunneling process being energy conserving. In this context, the value of $\mathcal{V}(\varphi_\mathcal{I})$ indicates the most probable tunneling height in the potential.
This point is determined by a combination of the classical probability of finding the particle at a specific height in the potential well at a specific temperature and the probability of tunneling through the barrier at that specific height. We schematically visualize this in Fig.~\ref{Fig.bouneexp}, where the overlap of the different colours indicates the regime of the most probable escape point. 
Hence, by increasing the temperature the particle is statistically elevated to higher points of the potential with a smaller barrier it tunnels through.

In the presence of dissipative couplings, the bounce path changes \cite{Hanggi1984}. Depending on whether the quantum fluctuations  and correspondingly the quantum decay rate  are/is enhanced or suppressed, the entry point $\mathcal{V}(\varphi_\mathcal{I})$ is shifted to higher/lower values, respectively. Furthermore, we find that in the presence of dissipation $\mathcal{V}(\varphi_\mathcal{I}) > \mathcal{V}(\varphi_\mathcal{E}) $ indicating that the tunneling event dissipates energy into the environment. We can study the most probable energy loss during the tunneling process by defining $E_d = \mathcal{V}(\varphi_\mathcal{I})-\mathcal{V}(\varphi_\mathcal{E})$ \cite{Weiss1984}. Hence, the interplay of the elevation of $\mathcal{V}(\varphi_\mathcal{I})$ and the energy loss  to the environment $E_d$ defines the exit point $\mathcal{V}(\varphi_\mathcal{E}) $. Depending on the temperature and the dissipative coupling the path can be above or below the non-dissipative one (see Fig.~\ref{Fig.EnResults}(a),(b) and (d),(f)).
Figure~\ref{Fig.EnResults} depicts the change of the above introduced quantities in the presence of either phase or charge dissipation. Thereby, we fix the parameters $I_C=21\, \mu$A, $C_{tot} = 6$ pF. Here, we set $C_J/C=1$ to show that also for a finite value of $C_J$ the impact of charge dissipation is still significant. 

\begin{figure}[t]
	\centering
    \includegraphics[scale=0.62]{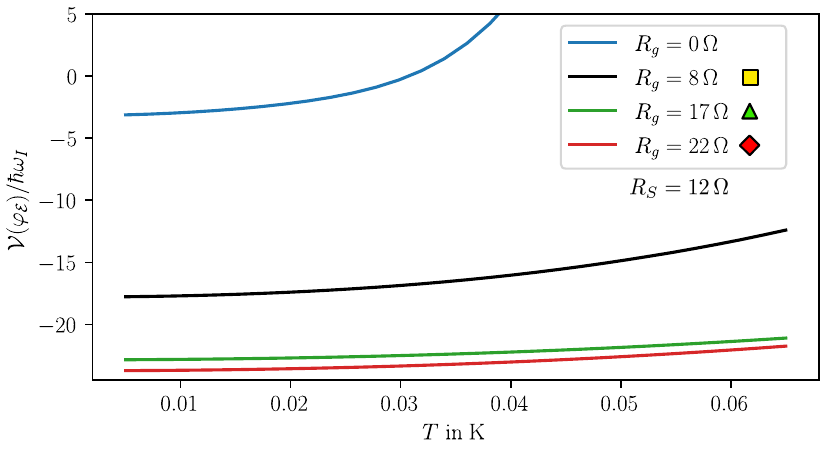}\\
	\caption{Escape points for $R_S=12\,\Omega$ and four different values of $R_g$ as a function of temperature. The presence of $R_g$ shifts the exit point significantly remaining independent of temperature over a broad range. System parameters: $I_C = 21\,\mu $A, $C_{tot}= 6$ pF, $I_b/I_C = 0.98$ and $C_J/C = 0.01$. }
	\label{Fig.TEscape}
\end{figure}

We summarize the conceptual difference between the impact of the two resistors in Fig.~\ref{Fig.EnResults}, where the top row contains the results for pure phase dissipation and the bottom row the ones for pure charge dissipation. Comparing (a) with (b), we find that the particle affected by pure charge dissipation (b) is tunneling below the non dissipative value irrespective of temperature, while for pure phase dissipation (a) the opposite happens. This is a consequence of the respective impact of the dissipative coupling on the decay rate. In the case of phase dissipation the quantum fluctuations are decreased and the particle has to be elevated to higher escape points it can effectively tunnel through. Therefore, the exponentially small thermal activation contributes to the tunneling and the rate becomes temperature dependent. For charge dissipation the quantum fluctuations are enhanced and the particle tunnels with a relatively high probability through broad barriers -- without the impact of the exponentially suppressed elevation to higher points by thermal fluctuations.
This interpretation is consistent with the results in Fig.~\ref{Fig.EnResults}c and d. There we show the decay rate as a function of the respective dissipative coupling for different temperatures. The symbols in the plots connect to the parameters of the  paths labelled in Figs.\ref{Fig.EnResults}a and b. Increasing $R_g$ (d), the tunneling rate is enhanced and does not strongly depend on temperature. For pure phase dissipation (c) and low temperatures, the decay is suppressed and increasing the temperature increases the rate significantly. Furthermore, we find that in this case also the functional dependence on $R_S$ changes for different temperatures. This is seen from the blue dashed line in (c) demonstrating an enhancement of the rate with increasing phase dissipation due the increasing impact of thermal fluctuations on the decay process. Hence,  at this temperature the environment does not squeeze phase fluctuations any more but induces thermal fluctuations.
Figs.~\ref{Fig.EnResults}e-f, show  the escape point as a function of the dissipative parameters. For charge dissipation (f) the escape points are always below the non-dissipative case irrespective of temperature. For phase dissipation,  a strong dependence on $T$ is found.  Specifically, for $T>35 $ mK an enhancement of $1/R_S$ yields a higher exit point. For lower temperatures,  dissipation into the environment dominates yielding lower exit points by increasing the dissipative coupling.  This highlights the subtle effects originating from the interplay between the statistical elevation to higher-lying entry points $\mathcal{V}(\varphi_\mathcal{I})$ induced by thermal fluctuations, the tunnel probability, and the dissipated energy. 
The respective plots of the dissipated energy as function of the resistors are shown in Figs.~\ref{Fig.EnResults}g and h. 
There, as a consequence of $C/C_J=1$ the energy loss saturates by increasing $R_g$ and is for small temperatures less compared to the phase coupling \footnote{To understand how this changes for $C/C_J=0.01$, see Fig.~\ref{Fig.TEscape}}. However, because  the $T_0$ is increased also $E_d$ is less affected by temperature (see Fig.~\ref{Fig.EnResults}h).

To also relate the tunneling problem to results obtained in Sec.~\ref{Sec.2}, in Fig.~\ref{Fig.TEscape}, we display the most probable exit points in the presence of both dissipative couplings in Fig.~\ref{Fig.colourstate}c. The system predominately escapes the barrier far below the metastable minimum over a broad temperature range. Combining Figs.~\ref{Fig.TEscape}, \ref{Fig.Rate}b and \ref{Fig.voltage} leads to a comprehensive picture of the phase dynamics in the presence of both dissipative couplings. The decay rate is enhanced (Fig.~\ref{Fig.Rate}b) while the exit point is far below the barrier top (Fig.~\ref{Fig.TEscape}) so that for $R_g = 17\,\Omega$ and $R_g = 22\,\Omega$ the dynamics is re-trapped in the adjacent potential minimum  with high probability (Fig.~\ref{Fig.17example}). Note that due to the low exit point in the quantum tunneling regime, the parameter regime of phase diffusion, as introduced in Sec.\ref{Sec.2} and discussed in Fig.~\ref{Fig.crossovertemp}, is even broader. Using the respective escape point $\varphi_\mathcal{E}$ as the initial condition for the equation of motion for the same circuit parameters as in Fig.~\ref{Fig.voltage}b, one finds that $\varphi_\mathcal{E}$ is beyond the lowest point shown in Fig.~\ref{Fig.17example}, indicating fast equilibration in the next potential minimum.
Hence, in the parameter regime close to the border between phase diffusion and the running state (see green triangle in Fig.~\ref{Fig.colourstate}c), the dynamics is strongly sensitive to the escape point $\varphi_\mathcal{E}$ and therefore also to the escape mechanism, i.e. thermal hopping or quantum tunneling. 

In summary we can conclude that the impact of charge dissipation makes the phase particle  to tunnel  predominantly through energetically lower portions of the potential barrier (broad barrier), however, with an increased tunneling rate combined with energy transfer into the environment from the system. Hence, such an environment can induce cooling processes. We will discuss such implications in the next section. 

%
%
%
%
%
%

\section{Potential applications} \label{Sec.Outlook}
Current biased Josephson systems have been proposed to be suitable detectors for single microwave photons. Thereby, the phase particle is trapped in a metastable minimum of the potential, where an incoming microwave photon can elevate the phase particle above the barrier yielding a running state in the washboard potential and therefore a detectable voltage signal.
Such a scenario has the drawback that after measuring a voltage state, the system has to be reinitialized into the metastable state to be able to detect a second microwave photon. 
A suitably engineered dissipative environment as proposed in this article could be beneficial in the sense that by increasing $R_g$ re-trapping in the next local minimum is accompanied by a voltage pulse with an amplitude that is of the order of the running state signal.
Furthermore, the dark count rate of the system can be suppressed by several orders of magnitude via decreasing the shunt resistance $R_S$. 
Now, assuming the phase particle is elevated above the barrier by a microwave photon, the classical dissipative dynamics of the systems predicts an acceleration of the phase dynamics due to a competition between $R_g$ and $R_S$, which is followed by a strongly damped behaviour close to a valley in the washboard potential yielding efficient re-trapping and therefore direct reinitialization (see Fig.~\ref{Fig.voltage}c).

\begin{figure}[t]
	\centering
    \includegraphics[scale=0.5]{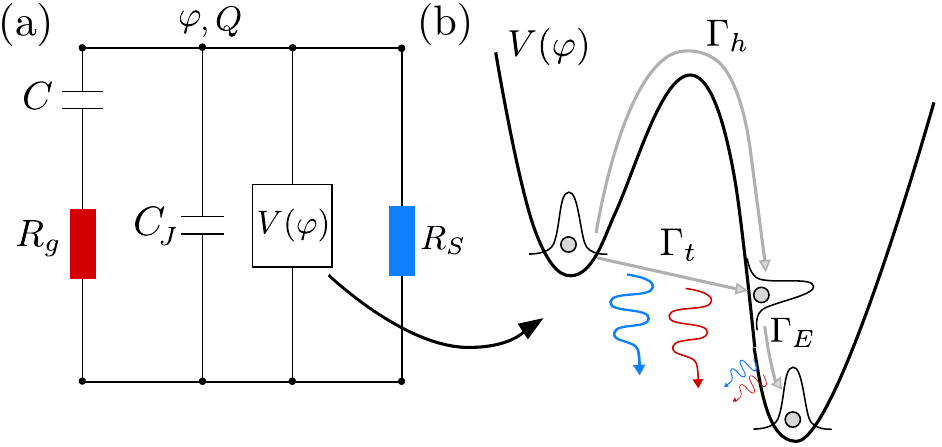}
	\caption{Situation of a phase particle trapped in a tilted double well potential. The results of this article can be used to qualitatively understand the behaviour in this and more general potentials.  }
	\label{Fig.annealing}
\end{figure}

Furthermore, the results of this article can easily be generalized to a variety of circuitQED experiments. Thereby, we can rely on the form of the Langevin equation (\ref{Eq.eqofmo}) and the equation determining the bounce trajectory Eq.~(\ref{Eq.EOM}). Using these equations, allows to  qualitatively study the situation displayed in Fig.~\ref{Fig.annealing},  by assuming a potential for which, after escaping the metastable potential well, the particles resides in another well on the right of the barrier. Because we choose the dissipative couplings to be strong, we can investigate this problem in a quasiclassical picture, where we assume that the phase is either tunneling incoherently into the well beyond or thermally hopping over the barrier.
The phase is therefore not in a coherent superposition between two wells but can be treated as a quasiclassical particle. As discussed in Sec.~\ref{Sec.meta}, depending on the dissipative couplings the particle escapes on different time scales and also tunnels at different positions.
The result that increasing $R_g$ yields a significantly increased decay rate can be leveraged for dissipative accelerated cooling to a classical groundstate. 
Please note that, within the described quasiclassical context, the particles dynamics beyond the barrier is sufficiently described by the Langevin equation containing both resistors $R_g$ and $R_S$.
Making the assumption that the well on the right of the barrier is described by a harmonic potential, both dissipative couplings yield damped dynamics and fast equilibration to next minimum. 
The timescale of this procedure can be quantitatively compared to a preparation of the classical ground state via a non-dissipative adiabatic quantum process by relating the incoherent tunneling rate $\Gamma_t$ to the coherent tunnel splitting $\Delta$ between the wells. $\Delta$ is the time limiting factor during an adiabatic Landau Zener type evolution preparing the localized state in the right well of the potential. Assuming that the barrier does not change significantly during the evolution one can approximately relate $\Gamma_t \propto \Delta^2$. However, while the coherent tunnel splitting is fixed solely by the potential barrier, we showed that $\Gamma_t$ can be increased by several orders of magnitude by the anomalous coupling. For specific (especially for narrow) potentials the dissipative incoherent decay might be faster than the adiabatic process. 
Considering further  wells on the right, the process would repeat, eventually creating a funnel to the classical ground state of the potential. This process can be combined with thermal annealing (TA) procedures, where quantum tunneling could bring the particle from one local minimum that was found via TA to another one with smaller energy. Hence, at small enough temperature (at the end of the thermal annealing procedure) the dissipative enhanced incoherent quantum tunneling rate can be a resource for efficient cooling. This can be particularly interesting for complex optimization problems with multiple parameters. Similar ideas of utilizing dissipation to accelerate annealing procedures have  been proposed in \cite{Cirac2009,Smelyanskiy2017,Venuti2017,Arceci2018,Mishra2018,Salatino2025,Sveistrys2025,Najafabadi2023,zhou2026}. 

Furthermore, the impact of charge dissipation can become very rich provided the potential is parametrically driven. This can easily be seen by inserting a time dependent potential $\mathcal{V}(\varphi,t)$ into Eq.~(\ref{Eq.eqofmo}). The effect of momentum dissipation on driven systems was recently investigated in \cite{wagner2026}.

\section{Conclusion}\label{Sec.Conclusion}

In this article, we presented a detailed analysis of the impact of a $C-R_g$ shunt as defined in Fig.~\ref{Fig.1} and Fig.~\ref{Fig.annealing} on the dynamics of a phase particle in an anharmonic potential. Specifically, we showed that such a circuitry leads to anomalous dissipation, which in the quasiclassical limit is described by the Langevin equation Eq.~(\ref{Eq.eqofmo}).  This equation offers interesting insight into the impact of $R_g$ on the dynamics in any anharmonic potential. 
The current biased Josephson Junction studied in this work serves as an important example, where the impact of $R_g$ in combination with a shunt resistor $R_S$ leads to interesting classical and quantum effects. Importantly, we showed that the interplay between both resistors leads to re-trapping in the next minimum after a metastable decay from a local minimum in the potential. In this context, we also discussed the impact of retardation effects stemming from the kernel $\eta(t)$.

Considering the impact of the dissipative couplings on quantum effects, we studied the temperature dependence of the decay rate in the presence of both resistors. We found that, especially in the region of quantum tunneling, $R_g$ has a significant impact. The rate is enhanced by several orders of magnitude. In the parameter regime studied in this work, this happens already a low resistances. To gain more insight into the properties of the decay channel, we also studied the impact of both resistors on the the most probable tunnel trajectory and  find that charge coupling makes the particle predominantly tunnel deep in the well for a large temperature regime.  
Our results offer important insights and a new perspective on the dissipative quantum dynamics of the superconducting phase in non-linear potentials, with wide applicability. 

\acknowledgments
We are grateful to M. Devoret, C. Padurariu, C. Gatti, E. Pollak, D. Golubev, J. Stockburger, O. Zilberberg and G. Rastelli
for fruitful discussions.
%


\appendix

\section{Thermally Activated Decay}\label{sec:ThermallyActivatedDecay}
In the following section, we describe the escape out of the well in the regime where it is dominated by thermal activation. As explained in the main text, in the high temperature limit the imaginary time domain $\beta$ becomes small and the only possible (meta-)stable solutions for the extremal paths in the partition function Eq.~(\ref{Eq.partfunc})
come from the vicinity of the stationary points of the action, i.e. of the time independent trajectories $\varphi(\tau)=0$ and $\varphi(\tau)=\varphi_b$. Furthermore, the solutions for the most probable path $\varphi(\tau)$ fulfill the boundary condition $\varphi(-\beta/2)=\varphi(\beta/2)$. The point $\varphi(\tau)=0$ corresponds to the phase particle sitting on top of the inverted potential whereas the other trivial solution to the particle staying in the well of the inverted potential.
We approximate the potential with expansions up to second order and evaluate the action in the vicinity around the stationary points \cite{grabert_temperature_1985, Grabert1987}.

A periodic path in the vicinity of the time independent path is written as 
\begin{align}\label{eq:extremalPath}
\varphi(\tau)=\tilde{\varphi}+\varphi_t \Theta \sum_{n=-\infty}^\infty \Phi_n e^{i\omega_n\tau}
\end{align}
where we defined the scaling factor $\varphi_t=3\omega_I^2/(2\omega_J^2)$ from the condition $V(0)=V(\varphi_t)$ , $\Theta=2\pi k_B T/(\hbar\omega_I)$ the dimensionless temperature, $\omega_n$ to be the $n$-th Matsubara frequency and $\tilde{\varphi}=\{0, \varphi_b\}$ is the stationary path.

Around the stationary paths we perform a harmonic approximation of the equation of motion Eq.~(\ref{Eq.EOM}), solve it using the ansatz in Eq.~(\ref{eq:extremalPath}) and find the effectibe action around the stationary paths. For $\varphi(\tau)=0$ we obtain 
\begin{align}\label{eq:EffAcWellHighT}
S_\text{eff}^0=\frac{\Phi_0^2C_\text{tot}}{2}\beta\varphi_t^2\Theta^2\sum_{n=-\infty}^{\infty}\lambda_{n}^{0}\Phi_{n}\Phi_{-n}
\end{align}
where
\begin{align}\label{eq:LambdaN0}
    \lambda_n^{0}=\omega_{n}^{2}\left(1-\frac{1}{1+\frac{C_J}{C}}\frac{\tau_{p}\frac{|\omega_{n}|}{1+\frac{|\omega_{n}|}{\omega_{C}}}}{1+\tau_{p}\frac{|\omega_{n}|}{1+\frac{|\omega_{n}|}{\omega_{C}}}}\right)+\gamma\frac{|\omega_{n}|}{1+\frac{|\omega_{n}|}{\omega_{C}}}+\omega_{I}^{2}\,.
\end{align}
The contribution of the effective action in Eq.~(\ref{eq:EffAcWellHighT}) to the partition function ${Z}$ is evaluated by performing the Gaussian integrals over the amplitudes $\Phi_n$.
We deal with the integral measure of the path integral and use \cite{Weiss2012,kleinert_path_2006}
\begin{align}
    \oint\mathrm{D}[\varphi(\tau)]\to\mathcal{N}\prod_{n=-\infty}^{\infty}\int_{-\infty}^{\infty}\frac{\mathrm{d}\Phi_{n}}{\sqrt{2\pi\hbar}}
\end{align}
with the normalization constant $\mathcal{N}$.
Thus, the contribution to the partition function of the particle sitting on top of the barrier in the inverted potential is exactly evaluated and yields
\begin{align}\label{eq:PartitionFuncPotMinimum}
    {{Z}}_0=\mathcal{N}\sqrt{\frac{1}{{\Phi_0^2C_\text{tot}}\beta\varphi_{t}^{2}\Theta^{2}\lambda_{0}^0}}\prod_{n=1}^{\infty}\frac{1}{M\beta\varphi_{t}^{2}\Theta^{2}}\frac{1}{\lambda_n^0}\,.
\end{align}
The calculation for the other time independent path $\varphi(\tau)=\varphi_b$ is analogous.
The second order action reads
\begin{align}\label{eq:EffAcBarrierHighT}
    S_\text{eff}^b=\frac{1}{2}{\Phi_0^2C_\text{tot}}\varphi_t^2\Theta^2\beta\sum_{n=-\infty}^{\infty}\lambda_{n}^{b}\Phi_{n}\Phi_{-n}+\beta \mathcal{V}_0
\end{align}
where 
\begin{align}\label{eq:eigenvaluesLambdaB}
    \lambda_n^{b}=\omega_{n}^{2}\left(1-\frac{1}{1+\frac{C_J}{C}}\frac{\tau_{p}\frac{|\omega_{n}|}{1+\frac{|\omega_{n}|}{\omega_{C}}}}{1+\tau_{p}\frac{|\omega_{n}|}{1+\frac{|\omega_{n}|}{\omega_{C}}}}\right)+\gamma\frac{|\omega_{n}|}{1+\frac{|\omega_{n}|}{\omega_{C}}}-\omega_{I}^{2}\,.
\end{align}
We can already see in Eq.~(\ref{eq:EffAcBarrierHighT}) that we get an additional factor of $\beta \mathcal{V}_0$.
This factor results in the Arrhenius factor at high temperatures.
Further, we recognize that the eigenvalue $\lambda_0^b=-\omega_I^2$ is negative and thus, we cannot evaluate the Gaussian integral for the amplitude $X_0$ since it diverges.
This should not be a surprise since we calculate the free energy of an unstable system.
However, we still evaluate the integral using the method introduced by Langer \cite{langer_statistical_1969} and obtain
\begin{align} \label{eq:PartFunctionBarrierHighT}
    {{Z}}_B=\frac{i}{2}\mathcal{N}\sqrt{\frac{1}{{\Phi_0^2C_\text{tot}}\beta\varphi_{t}^{2}\Theta^{2}|\lambda_{0}^b|}}\prod_{n=1}^{\infty}\frac{1}{M\beta\varphi_{t}^{2}\Theta^{2}}\frac{1}{\lambda_n^b}e^{-\frac{\beta}{\hbar}\mathcal{V}_{0}}
\end{align}
where we explicitly see the imaginary part which is exponentially small due to the factor $\exp(-\mathcal{V}_0/(k_\text{B}T))$.
It is consistent within this approximation to keep the exponentially small expression since it is the leading imaginary part.
We notice that when lowering the temperature the eigenvalue $\lambda_1^b$ vanishes at a certain temperature which we refer to as crossover temperature $T_0$.
Thus, the partition function diverges for $T\to T_0$ but this divergence is not of a physical origin but rather the result of the method and Gaussian approximation employed for the fluctuations \cite{Weiss2012}. In this way, the crossover temperature defines the range of applicability of the harmonic approximation around the stationary solutions for thermally activated decay. How to mathematically avoid the unphysical divergence was recently discussed in \cite{pollak}.

The partition function of the entire problem is then the sum of the two defined parts above, which we write as ${Z}={Z}_0(1+{Z}_b/{Z}_0)$ where the second term is imaginary and contains an exponentially small factor.
The imaginary part of the free energy $F=-k_B T \ln ({Z})$ yields
\begin{align}
    \mathrm{Im}(F) &= -k_B T\, \mathrm{Im}\left[\ln\left({Z}_0\left(1+\frac{{Z}_B}{{Z}_0}\right)\right)\right]\notag\\
    &\approx -k_B T\,\frac{1}{{Z}_0}\mathrm{Im}({Z}_B)\notag\\
    &=-\frac{k_B T}{2}\sqrt{\frac{\lambda_{0}^{0}}{|\lambda_{0}^{b}|}}\prod_{n=1}^{\infty}\frac{\lambda_{n}^{0}}{\lambda_{n}^{b}}e^{-\frac{\mathcal{V}_0}{k_\text{B}T}}
\end{align}
and the escape rate is then given as 
\cite{langer_statistical_1969, Affleck1981}
\begin{align}
    \Gamma = -\frac{2}{\hbar}\frac{T_0}{T}\mathrm{Im}(F)
\end{align}
where we introduced an additional factor of $T_0/T$ which takes care of the effect of re-crossings at the barrier top.
We obtain
\begin{align}\label{eq:EscapeRateAboveT0}
    \Gamma = \underset{\Gamma_{cl}}{\underbrace{\frac{\omega_I}{2\pi}\Theta_0 e^{-\frac{\mathcal{V}_0}{k_\text{B}T}}}}\,\,\underset{f_{q}}{\underbrace{\prod_{n=1}^{\infty}\frac{\lambda_n^0}{\lambda_n^b}}}
\end{align}
where we defined the temperature $\Theta_0=2\pi k_\text{B} T_0/(\hbar\omega_I)$ and splitted the decay rate into a classical part $\Gamma_\text{cl}$ and a part containing the quantum correction factors $f_q$.

\section{Quantum Tunneling}\label{sec:QuantumTunneling}
In the temperature region below the crossover temperature $T_0$ quantum fluctuations are getting more and more important for the escape out of the metastability.
In this section, we first introduce the appropriate method for treating quantum tunneling below $T_0$.

\subsection{Bounce Solution}\label{sec:BounceSolution}
Below the crossover temperature $T_0$ an additional non-trivial periodic solution of the equation of motion (\ref{Eq.EOM}) emerges the bounce solution $\varphi_B(\tau)$.
In this temperature regime, the period $\beta$ is long enough to allow for this oscillating solution.
The bounce fulfils the boundary conditions $\varphi_B(0)=\varphi_B(\beta)$ and $\dot{\varphi}_B(0)=\dot{\varphi}_B(\beta)$ and describes the tunneling event.
We describe the oscillating bounce solution with the Fourier Ansatz \cite{grabert_temperature_1985}
\begin{align}\label{eq:BounceAnsatz}
    \varphi_B(\tau)=\varphi_t\Theta\sum_{n=-\infty}^\infty X_n\,e^{-i\omega_n\tau}
\end{align}
where $\varphi_t=3\omega_I^2/(2\omega_J^2)$ is obtained from the condition $V(0)=V(\varphi_t)$ and $\Theta = 2\pi k_\text{B}T/(\hbar\omega_I)$ is the dimensionless temperature.
When $\varphi_B(\tau)$ is an extremal solution of the equation of motion then also the path $\varphi_B(\tau+\tau_0)$ is an extremal action path.
However, this change of the phase of the bounce has no effect onto the numerical value of the action and thus, the bounce is translational invariant and has a zero mode, as we will see later \cite{Weiss2012}.
It is convenient to choose the bounce with the symmetry $\varphi_B(-\tau)=\varphi_B(\tau)$ leading to the condition $X_n^\star=X_{-n}$ for the Fourier coefficients.

We insert the ansatz Eq.~ (\ref{eq:BounceAnsatz}) into the equation of motion (\ref{Eq.EOM}) yielding for the Fourier coefficients 
\begin{align}\label{eq:BounceFourierCoefficients}
    \lambda_nX_n=\frac{3}{2}\Theta\left(2\sum_{k=1}^{\infty}X_{n+k}X_{k}+\sum_{k=0}^{n}X_{n-k}X_{k}\right)
\end{align}
where 
\begin{align}\label{eq:eigenvaluesLambdaBelow}
    \lambda_n =&  \nonumber n^{2}\Theta^{2}\left(1-\frac{1}{1+\frac{C_{J}}{C}}\frac{\tau_{p}\omega_{I}\frac{|n|\Theta}{1+\frac{|n|\Theta\omega_{I}}{\omega_{C}}}}{1+\tau_{p}\omega_{I}\frac{|n|\Theta}{1+\frac{|n|\Theta\omega_{I}}{\omega_{C}}}}\right)\\ &+\frac{\gamma}{\omega_{I}}\frac{|n|\Theta}{1+\frac{|n|\Theta\omega_{I}}{\omega_{C}}}+1\,.
\end{align}
Using Eq.~(\ref{eq:BounceFourierCoefficients}), we solve for the bounce Fourier coefficients in an numerical manner.
\begin{figure}[t]
        \centering
\includegraphics[height=15cm]{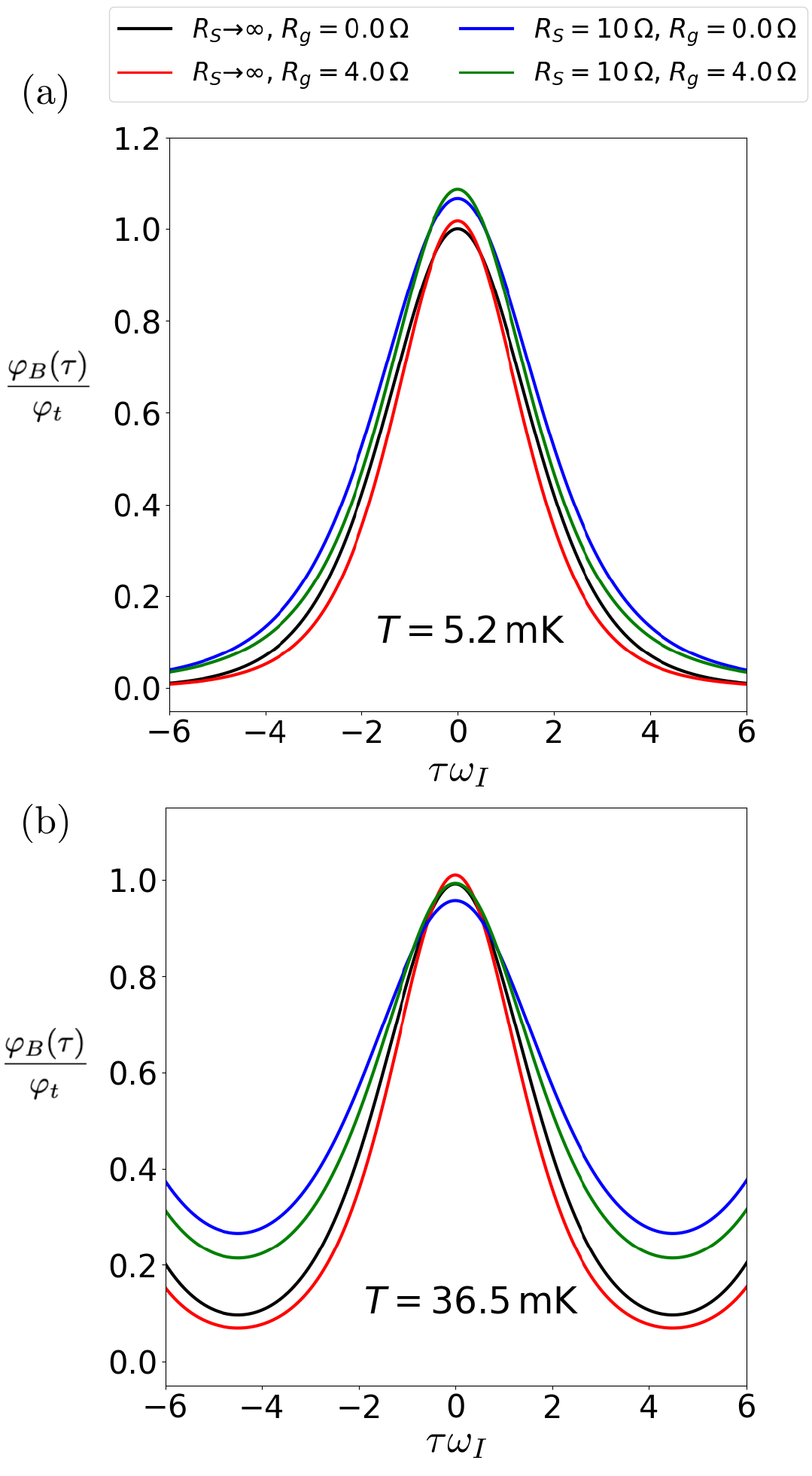}   
    \caption{(a) Bounce solution for different coupling strengths at the temperature $5.2\,$mK and (b) $36.5\,$mK normalized with the non-dissipative tunneling length $\varphi_t$ against the imaginary time $\tau$. Dissipation changes the shape, the period and the penetration depth (the amplitude) of the bounce.}
    \label{fig:BounceSolution}
\end{figure}
We follow the iterative procedure introduced first by Chang and Chakravarty \cite{chang_quantum_1984} for determination of the zero temperature rate and an extension to finite temperatures $0.1\Theta_0<\Theta<0.95\Theta_0$ from Grabert et. al \cite{grabert_temperature_1985, Grabert1987}.
The algorithm developed by Chang and Chakravarty avoids a dangerous direction in the straightforward iteration.
They modified (\ref{eq:BounceFourierCoefficients}) by introducing the parameter $\mu$ to make the iteration stable and get
\begin{align}\label{eq:IterationBounce}
    X_n=\frac{1}{\lambda_n}\frac{3}{2}\Theta\mu\left(2\sum_{k=1}^{\infty}X_{n+k}X_{k}+\sum_{k=0}^{n}X_{n-k}X_{k}\right)\,.
\end{align}
The algorithm consists of the following steps.
\begin{enumerate}
    \item Start by initializing the coefficients $X_n^{(0)}$ (the superscript denotes the iteration step) with a zeroth-order approximation ${X_n^{(0)}\propto\,e^{-n}}$. Also initialize $\mu^{(0)}=1$

    \item Perform the iteration step $X_n^{(1)}$ with the right hand side of Eq.~\ref{eq:IterationBounce} using the $X_n^{(0)}$ and $\mu^{(0)}$.

    \item Update the parameter $\mu^{(1)} = \mu^{(0)} \left(\frac{X_0^{(0)}}{X_0^{(1)}}\right)^2$.

    \item Find the next iteration step $X_n^{(2)}$.

    \item Repeat the steps 2-4 until the coefficients converges, i.e. the difference between the steps satisfies a predefined criterion.
\end{enumerate}
We observe convergence after a few tens of steps over the whole parameter and temperature range.
The Fourier coefficients $X_n$ decay rapidly for increasing $n$'s and we truncate the sum in Eq.~(\ref{eq:BounceFourierCoefficients}) at $N=500$ since we do not observe any changes for larger $N$.

We show the obtained solutions for the bounce in Fig.~\ref{fig:BounceSolution} for the temperatures ${T=5.2\,\text{mK}}$ (corresponds to $\Theta_0=0.1$ for the non dissipative case) and ${T = 36.5\,\text{mK}}$ (corresponds to $\Theta_0=0.7$ for the non dissipative case) in Fig.~\ref{fig:BounceSolution} for different coupling strengths to the phase and the charge, respectively.
As a consequence of alterations in temperature, the period $\beta$ of the bounces shortens. Furthermore, the amplitude of the bounces which corresponds to the average tunneling length gets smaller.
Further, the amplitude of the bounces also changes with dissipation.
Coupling to the phase shortens the amplitude and shifts the bounce upward whereas coupling to the charge enlarges the amplitude and shifts it downward.
The bounce oscillates around the minimum of the inverted potential with the height of the oscillation decreasing as the  temperature increases.
In the presence of dissipation, the reversal point is higher than the starting point corresponding to energy dissipation.

\subsection{Escape Rate Below $T_0$} \label{Sec:Below}
Below $T_0$, there are three saddle points of the action, the two trivial solutions ${\varphi(\tau)=0}$ and $\varphi(\tau)=\varphi_b$ and the bounce solution $\varphi_B(\tau)$.
Since the action of the bounce is significantly smaller than the action $\beta \mathcal{V}_0$ of the trivial solution $\varphi(\tau)=\varphi_b$, the bounce is the leading contribution of the barrier to the partition function and to the imaginary part of the free energy.
We neglect the trivial solution $\varphi_b$ therefore in the following \cite{Weiss2012}.

In the semiclassical approximation the main contribution to the partition function originates from paths in the vicinity of the bounce saddle point.
We use the Ansatz $\varphi(\tau)=\varphi_B(\tau)+\xi(\tau)$ and obtain the second order action
\begin{align}
    S[\varphi(\tau)]
    &=S_B+\frac{1}{2}{\Phi_0^2C_\text{tot}}\int_0^\beta\mathrm{d}\tau \xi(\tau) L_B[\xi(\tau)]
\end{align}
where $\xi(\tau)$ are fluctuations around the bounce, and we also defined the linear operator $L_B[\xi(\tau)]$ which is given by
\begin{align}\label{eq:LinearOperator}
    L_B[\xi(\tau)]&=-\ddot{\xi} \nonumber(\tau)+\frac{1}{{\Phi_0^2C_\text{tot}}}V''(\varphi_{B}(\tau))\xi(\tau)\\&+\frac{1}{{\Phi_0^2C_\text{tot}}}\int_{0}^{\beta}\mathrm{d}\tau'F^{(\varphi)}(\tau-\tau')\xi(\tau')\nonumber\\
    &-\frac{1}{{\Phi_0^2C_\text{tot}}}\int_{0}^{\beta}\mathrm{d}\tau'F^{(Q)}(\tau-\tau')\ddot{\xi}(\tau')
\end{align}
The stationary bounce action can readily be evaluated in terms of Fourier coefficients of the bounce and reads in terms of the Fourier coefficients $X_n$
\begin{align}
    S_B= 9\pi\hbar v \Theta\Big(\frac{1}{2}X_{0}^{2}+\sum_{n=1}^{\infty}\lambda_n X_{n}^{2}\Big)\,.
\end{align}
To evaluate the effect of the fluctuations around the bounce trajectory, we chose a periodic Ansatz that reads in Fourier representation \cite{grabert_temperature_1985}
\begin{align}\label{eq:AnsatzFlucBounce}
    \xi(\tau)=\varphi_t\Theta\left(Y_0+\sqrt{2}\sum_{n=1}^\infty\left( Y_n\cos(\omega_n\tau)+\mathcal{Z}_n\sin(\omega_n\tau)\right)\!\!\right)
\end{align}
where the $\omega_n$ are the Matsubara frequencies.
We split the Ansatz in Eq.~(\ref{eq:AnsatzFlucBounce}) into an even part $\xi_\text{even}(\tau)$ (the coefficients $Y_l$) and an odd part $\xi_\text{odd}$ (the coefficients $Z_l$).
The even and odd part of the Ansatz are eigenfunctions of the linear operator in Eq.~(\ref{eq:LinearOperator}).
Therefore, we define the eigenvalue problems $L_B[\xi_\text{even}]=b\omega_I^2\xi_\text{even}$ and ${L_B[\xi_\text{odd}]=a\omega_I^2\xi_\text{odd}}$ with the dimensionless eigenvalues $a$ and $b$ \cite{Grabert1987}.
We solve this eigenvalue problem
and obtain the matrices $A_{n,m}$ and $B_{n,m}$ with eigenvalues $a_n$ (with $n=0,1,2,...$) and $b_n$ (with $n=1,2,...$), respectively.
The matrix elements are determined the Eqs.~(\ref{eq:BounceAnsatz}) and (\ref{eq:LinearOperator}) yielding
\begin{align}\label{eq:matricesAB}
    A_{00}&=1-3\Theta X_{0}\nonumber\\
    A_{n0}&=-3\sqrt{2}\Theta X_{n}\nonumber\\
    A_{0n}&=-3\sqrt{2}\Theta X_{n}\nonumber\\
    A_{nm}&=\lambda_{n}\delta_{n,m}-3\Theta(X_{|n-m|}+X_{n+m})\nonumber\\
    B_{nm}&=\lambda_{n}\delta_{n,m}-3\Theta(X_{|n-m|}-X_{n+m})
\end{align}
where the $X_n$ are the Fourier coefficients of the bounce ansatz.
With the dimensionless eigenvalues of these matrices, the second order action correction for the fluctuations reads
\begin{align}
    \Delta S&=\frac{27}{2}\hbar v\Theta\pi\Big(\sum_{n=0}^{\infty}a_{n}Y_{n}Y_{n}+\sum_{n=1}^{\infty}b_{n}\mathcal{Z}_{n}\mathcal{Z}_{n}\Big)\,.
\end{align}
Here, we encounter again the problem of a negative eigenvalue and of a vanishing eigenvalue since the eigenvalue $a_0$ is negative and the eigenvalue $b_1$ vanishes.
The negative eigenvalue points to the fact that the system is unstable and the vanishing eigenvalue takes account of the translational invariance of the bounce solution, as already mentioned.
We introduce the integral measure to integrate out the amplitudes $Y_n$ and $Z_n$ reading \cite{Weiss2012, kleinert_path_2006}
\begin{align}
    \oint\mathrm{D}[\varphi(\tau)]\to{\cal N}\prod_{n=0}^{\infty}\prod_{m=1}^{\infty}\int_{-\infty}^{\infty}\frac{\mathrm{d}Y_{n}}{\sqrt{2\pi\hbar}}\int_{-\infty}^{\infty}\frac{\mathrm{d}\mathcal{Z}_{m}}{\sqrt{2\pi\hbar}}\,.
\end{align}
The exact integration along the steepest descent 
yields
\begin{align}
    {Z}_B&=\frac{i}{2}\mathcal{N}\beta\frac{e^{-\frac{1}{\hbar}S_{B}}}{\sqrt{{\Phi_0^2C_\text{tot}}\beta\varphi_{t}^{2}\Theta^{2}}}\\& \nonumber \times \frac{\sqrt{S_{0}}}{\sqrt{2\pi\hbar}}\frac{1}{{\Phi_0^2C_\text{tot}}\beta\varphi_{t}^{2}\Theta^{2}\omega_{I}^{2}}\sqrt{\frac{1}{|a_{0}|a_{1}}}\\& \nonumber \times \prod_{n=2}^{\infty}\frac{1}{{\Phi_0^2C_\text{tot}}\beta\varphi_{t}^{2}\Theta^{2}\omega_{I}^{2}}\sqrt{\frac{1}{a_{n}b_{n}}} \label{Eq.FB}
\end{align}
where the eigenvalue $a_0$ is negative and the eigenvalue $b_1$ vanishes.
The factor ${S_0={\Phi_0^2C_\text{tot}}\int_{-\beta/2}^{\beta/2}\mathrm{d}\tau\dot{\varphi}^2_B(\tau)}$ is the zero-mode normalization factor which occurs by the change of the integration variable.
The partition function for the contribution of the stationary point at the potential minimum is given in \ref{eq:PartitionFuncPotMinimum} with the eigenvalues $\lambda$ defined in \ref{eq:eigenvaluesLambdaBelow}.
The partition function of the escape problem below the crossover temperature $T_0$ is then the sum of the two partition functions above, which is written as ${Z}={Z}_0(1+{Z}_B/{Z}_0)$.
The term originating from the bounce solution contains the imaginary part which is exponentially small.
The imaginary part of the free energy evaluates as
\begin{align}
    \mathrm{Im}(F)&\approx-k_B T\,\frac{1}{{Z}_0}\mathrm{Im}({Z}_B)\\
    &=-k_B T\,\frac{\beta\omega_{I}}{2}\sqrt{\frac{S_{0}}{2\pi\hbar}}\frac{\lambda_{1}e^{-\frac{1}{\hbar}S_{B}}}{\sqrt{|a_{0}|a_{1}}}\prod_{n=2}^{\infty}\frac{\lambda_{n}}{\sqrt{a_{n}b_{n}}}.\nonumber
\end{align}
To determine the escape rate, we use $\Gamma=-2/\hbar\,\mathrm{Im}(F)$ \cite{langer_statistical_1969} and get for the particle escaping out of the metastable state for temperatures below the crossover temperature 
\begin{align}\label{eq:EscapeRateBelowT0}
    \Gamma&=\frac{\omega_I}{2\pi}\chi e^{-\frac{1}{\hbar}S_B}
\end{align}
where we introduced the dimensionless prefactor
\begin{align}
    \chi=2\pi\sqrt{g}\frac{\lambda_{1}}{\sqrt{|a_{0}|a_{1}}}\prod_{n=2}^{\infty}\frac{\lambda_{n}}{\sqrt{a_{n}b_{n}}}
\end{align}
with $g=27\Theta\sum_{n=0}^{\infty}n^{2}\Theta^{2}X_{n}^{2}$ from the zero-mode normalization factor $S_0$.

To determine the pre-exponential factor of the escape rate below the temperature $T_0$ in Eq.~(\ref{eq:EscapeRateBelowT0}), we have to calculate the eigenvalues $a_n$ and $b_n$ of the matrices $A_{n,m}$ and $B_{n,m}$ in Eq.~(\ref{eq:matricesAB}) which is difficult since the matrices are of infinite dimension.
Therefore, we have to apply some approximations, which are presented in the following:
The eigenvalues $a_n$ and $b_n$ approach $\lambda_n$ for large $n$ since the Fourier coefficients $X_n$ decay rapidly in this limit.
Therefore, for large $n$ the eigenvalues are calculated perturbative.
This yields \cite{grabert_temperature_1985}
\begin{align}\label{eq:PerturbativeEigenvalues}
    a_n&=\!\lambda_n\!-\!3\Theta(X_0\!+\!X_{2n})+9\Theta^2\sum_{m=0}^\infty{\vphantom{\sum}}'\,\frac{(X_{|n-m|}\!+\!X_{n+m})^2}{\lambda_n-\lambda_m}\nonumber\\
    b_n&=\!\lambda_n\!-\!3\Theta(X_0\!-\!X_{2n})+9\Theta^2\sum_{m=0}^\infty{\vphantom{\sum}}'\,\frac{(X_{|n-m|}-\!X\!_{n+m})^2}{\lambda_n-\lambda_m}
\end{align}
where $\sum_m^{'}$ sums over all values except for $m=n$.
We apply the following scheme for calculating the eigenvalues $a_n$ and $b_n$.
For small $n$, the eigenvalues are calculated by diagonalizing the truncated $N\times N$ matrices $A_{n,m}$ and $B_{n,m}$.
We choose $N$ such that at least ten consecutive eigenvalues agree with the perturbative result from Eq.~(\ref{eq:PerturbativeEigenvalues}) within an error of less than $10^{-6}$.
For even larger $n$'s, we use the approximation
\begin{align}
    a_n=b_n=\lambda_n-3\Theta X_0
\end{align}
for the eigenvalues.
Within these approximations, the product in the prefactor $\chi$ in Eq.~(\ref{eq:EscapeRateBelowT0}) converges within $10^6$  iterations over the whole parameter range explored.
We abort the iterations if the product does not change more than $10^{-2}$ within 10000 iterations.

\bibliographystyle{mybibstyle}
 
\bibliography{bibliography}

\end{document}